\documentclass[a4paper,10pt,twoside,bibnote]{cpc-hepnp}
\usepackage{CJK,upgreek,fancyhdr}
\usepackage{graphicx}
\usepackage{booktabs}
\usepackage{slashed}
\usepackage{amssymb,bm,mathrsfs,bbm,amscd}
\usepackage[tbtags]{amsmath}
\usepackage{lastpage}
\usepackage{multirow}
\usepackage{textcomp}
\usepackage{epstopdf}
\usepackage{float}
\usepackage{footnote}
\usepackage{CJK}
\usepackage{fancyhdr}
\begin{document}
\begin{CJK*}{GBK}{song}

\setlength{\abovecaptionskip}{4pt plus1pt minus1pt}
\setlength{\belowcaptionskip}{4pt plus1pt minus1pt}
\setlength{\abovedisplayskip}{6pt plus1pt minus1pt}
\setlength{\belowdisplayskip}{6pt plus1pt minus1pt}
\addtolength{\thinmuskip}{-1mu}
\addtolength{\medmuskip}{-2mu}
\addtolength{\thickmuskip}{-2mu}
\setlength{\belowrulesep}{0pt}
\setlength{\aboverulesep}{0pt}
\setlength{\arraycolsep}{2pt}


\title{A discussion on vacuum polarization correction to the cross-section of $\bm{e^+e^-\to\gamma^\ast/\psi\to\mu^+\mu^-}$}

\author{
      Hong-Dou Jin$^{1,2}$%
\quad Li-Peng Zhou$^{1,2;1)}$\email{zhoulipeng@ihep.ac.cn}%
\quad Bing-Xin Zhang$^{2}$%
\quad Hai-Ming Hu$^{2,1;2)}$\email{huhm@ihep.ac.cn}%
}
\maketitle

\address{
$^1$ University of Chinese Academy of Science, Beijing 100049, China\\
$^2$ Institute of High Energy Physics, Chinese Academy of Sciences, Beijing 100049, China\\
}

\begin{abstract}
Vacuum polarization is a part of the initial-state radiative correction for the cross-section of $e^+e^-$ annihilation processes. In the energy region in the vicinity of narrow resonances $J/\psi$ and $\psi(3686)$, the vacuum polarization contribution from the resonant component has a significant effect on the line-shape of the lepton pair production cross-section. This paper discusses some basic concepts and describes an analytical calculation of the cross-section of $e^+e^-\to\gamma^\ast/\psi\to\mu^+\mu^-$ considering the single and double vacuum polarization effect of the virtual photon propagator. Moreover, it presents some numerical comparisons with the traditional treatments.
\end{abstract}

\begin{keyword}
electron--positron annihilation, resonance, vacuum polarization, cross-section
\end{keyword}

\begin{pacs}
13.66.-a, 13.20.Gd, 12.15.Lk     \qquad     {\bf DOI:} 10.1088/1674-1137/43/1/013104
\end{pacs}


\vspace{2.1mm}

\section{Introduction}

In quantum field theory, tree-level Feynman diagrams represent a basic process of elementary particles reaction from the initial state to the final state, and the corresponding lowest order cross-section with order $\alpha^2$ is called Born cross-section. For accurate calculation, the contribution of higher level Feynman diagrams needs to be considered.

Among all the reactions in $e^+e^-$ annihilation, $e^+e^-\to e^+e^-$ and $\mu^+\mu^-$ are the two simplest quantum electrodynamics (QED) processes. Calculations of the unpolarized $e^+e^-\to e^+e^-$ and $\mu^+\mu^-$ cross-sections to order $\alpha^3$ (${\cal O}(\alpha)\sim1\%$) correction were studied decades ago\cite{npb571973381, npb631973381, npb681974541, npb1151976114}. Typically, radiative correction includes vertex correction, electron self-energy, vacuum polarization (virtual photon self-energy), and bremsstrahlung \cite{peskin}.

For perturbative calculations up to order $\alpha^3$, the radiative correction terms are the interferences between the tree level and higher level (one-loop) Feynman diagrams. In the references mentioned above, all the radiative correction terms were treated as small quantities owing to the extra factor, $\alpha$, compared to that in the tree-level terms. Such approximations for the QED correction and non-resonant quantum chromodynamics (QCD) hadronic correction are reasonable. However, for the energy regions in the vicinity of narrow resonances, such as charmonium $J/\psi$ and $\psi(3686)$, the contribution of the resonant component of the vacuum polarization (VP) correction is neither a small quantity nor a smooth function of energy. This implies that the energy dependence of the VP correction factor has a significant influence on the line shape of the total cross-section. Therefore, the VP correction in the vicinity of narrow resonances has to be treated appropriately.

The radiative correction of process $e^+e^-\to\mu^+\mu^-$ includes the initial-state and final-state corrections. The final-state radiative (FSR) correction is much smaller than the initial-state radiative (ISR) correction owing to the mass relation, $m_e\ll m_\mu$\cite{jackson}. The FSR correction can be neglected if one dose not require very high accuracy. In addition, the contributions of the two-photon-exchange diagrams and asymmetry of $e^\pm$ and $\mu^\pm$ are less important. In this work, only the ISR correction of the process, $e^+e^-\to\mu^+\mu^-$, is considered to keep the discussion succinct, and the discussions only concentrate on the VP correction. The calculations for other correction terms follow the expressions given in the related references\cite{slac4160,slac5160}.

The calculations of the resonant cross-section and VP correction need the bare value of the electron width of the resonance, but the value cited in the particle data group (PDG)  is the experimental electron width, which absorbs the VP effect\cite{ystsai,agshamov}. Therefore, another motivation of this work is attempt to provide a scheme for extracting the bare electron widths of resonances $J/\psi$ and $\psi(3686)$ by fitting the measured cross-section of $e^+e^-\to\mu^+\mu^-$ and then obtain the value of the Born-level Breit--Wigner cross-section.

The basic properties of a resonance with $J^{PC}=1^{--}$ is characterized by its three bare parameters: nominal mass $M$, electron width $\varGamma_e$, and total width $\varGamma$. The values of the resonant parameters can be predicted by the potential model\cite{highercharmonia}, but the theoretical uncertainties are difficult to estimate. A reliable method for obtaining accurate values of the resonant parameters is to fit the measured leptonic cross section\cite{plb6852010134,anashin} or hadronic cross section\cite{plb6602008315} in the vicinity of these resonances. Extracting the bare values from experimental data can provide useful information to decide the theories and models.

The bare values of the resonant parameters are the input quantities for the calculation of ISR factor $1+\delta(s)$ in the measurement of the $R$ value, which is defined as the lowest level hadronic cross-section normalized by the theoretical $\mu^+\mu^-$ production cross-section in $e^+e^-$ annihilation\cite{besiiisr,besiir2004}. In fact, the total hadronic cross-section is measured with the experimental data:
\begin{equation}
\sigma_{ex}^{\rm tot}(s)=\frac{N_{\rm had}}{L\epsilon},\label{croxttotdef}
\end{equation}
where $N_{\rm had}$ is the number of hadronic events, $L$ is the integrated luminosity of the data samples, $\epsilon$ is the detection efficiency for $e^+e^-\to$ hadrons determined by the Monte Carlo method, and $s$ is the square of the center-of-mass energy of initial state $e^+e^-$. However, the quantity of interest in physics is Born cross-section $\sigma_{ex}^{0}(s)$, which is related to $\sigma_{ex}^{\rm tot}(s)$ by ISR factor $1+\delta(s)$ as follows:
\begin{equation}
\sigma_{ex}^{0}(s)=\frac{\sigma_{ex}^{\rm tot}(s)}{1+\delta(s)},\label{croxtborndef}
\end{equation}
and $R$ value is measured:
\begin{equation}
R=\frac{\sigma_{ex}^{0}(s)}{\sigma_{\mu\mu}^0(s)}=\frac{N_{\rm had}}{\sigma_{\mu\mu}^0L\epsilon[1+\delta(s)]},~~~~\sigma_{\mu\mu}^0(s)=\frac{4\pi\alpha^2}{3s}.\label{rdefinition}
\end{equation}
ISR factor $1+\delta(s)$ indicates the fraction of all the high-order Feynman diagram contributions to the Born cross-section, which is a theoretical quantity by definition:
\begin{equation}
1+\delta(s)\equiv\frac{\sigma^{\rm tot}(s)}{\sigma^{0}(s)},\label{isrfactordef}
\end{equation}
where $\sigma^{0}(s)$ and $\sigma^{\rm tot}(s)$ are the theoretical Born cross-section and total cross-section, respectively. The accurate calculation of $1+\delta(s)$ is a key factor for obtaining the $R$ value from the measured $\sigma_{ex}^{\rm tot}(s)$. The calculation of $\sigma^{\rm tot}(s)$ needs the values of $\sigma^0(s')$ from $s'=4m_{\pi}^2$ to $s$ as inputs. If the correlation between the continuum and resonant states can be neglected, the hadronic Born cross-section can be written as:
\begin{equation}
\sigma^0(s)=\sigma_{\rm con}^0(s)+\sigma_{\rm res}^0(s),\label{sigmaconres}
\end{equation}
where $\sigma_{\rm con}^0(s)=\sigma_{\mu\mu}^0(s)\tilde{R}(s)$, $\tilde{R}(s)$ is the $R$ value from which the resonant contribution has been subtracted. Generally, the Born-level resonant cross-section is expressed in the Breit--Wigner form:
\begin{equation}\label{bwformula}
\sigma_{\rm res}^0(s)=\frac{12\pi\varGamma_e\varGamma}{(s-M^2)^2+M^2\varGamma^2},
\end{equation}
where the resonant parameters $(M,\varGamma_e,\varGamma)$ must be bare quantities. The value of the electron width cited in the PDG is, in fact, the experimental value of $\varGamma_e^{ex}$, which absorbs the VP effect, but uses the same notation, $\varGamma_e$, as the bare one. If the users directly use the dressed value of $\varGamma_e^{ex}$ as the bare one, $\varGamma_e$, in Eq.~(\ref{bwformula}), then the value of $1+\delta(s)$ calculated by Eq.~(\ref{isrfactordef}) is incorrect. In this regard,
\begin{equation}\label{bwformulawrong}
\tilde{\sigma}_{\rm res}(s)=\frac{12\pi\varGamma_e^{ex}\varGamma}{(s-M^2)^2+M^2\varGamma^2}\neq\sigma_{\rm res}^0(s),
\end{equation}
and
\begin{equation}\label{isrdefinitionwrong}
\frac{\sigma^{\rm tot}(s)}{\sigma_{\rm con}^0(s)+\tilde{\sigma}_{\rm res}(s)}\neq1+\delta(s).
\end{equation}
Obviously, the obtained value from the left-hand-side of Eq.~(\ref{isrdefinitionwrong}) is VP double deducted. Even if a user notices that the $\varGamma_e^{ex}$ cited in the PDG is a dressed value, he does not know how to extract the bare value, $\varGamma_e$, from $\varGamma_e^{ex}$. If a user uses the value of $\varGamma_e$ predicted by the theoretical model, then it becomes difficult to control the uncertainty of $\varGamma_e$. Some models, for example, the potential model introduced in reference\cite{highercharmonia}, do not provide the theoretical uncertainty of $\varGamma_e$. Therefore, extracting $\varGamma_e$ from the data is necessary for the $R$ value measurement.

The discussion in the following sections will be concentrated on the VP correction of $\sigma^{\rm tot}(s)$ for the process, $e^+e^-\to\mu^+\mu^-$. The outline of this paper is as follows: In section 2, the related Born cross-sections are presented. In section 3, the VP correction to the virtual photon propagator described in text books and references is reviewed. In section 4, the experimental lepton width with different conventions is reviewed. In section 5, the properties of the VP-modified Born cross-section are discussed and the line-shapes are shown graphically. In section 6, the analytical expressions of the total cross-section of $e^+e^-\to\mu^+\mu^-$ with single and double VP corrections are deduced, and the numerical results are presented. Section 7 presents some discussions and comments.

\section{Born cross-section}

In the energy region containing resonance $\psi$, final state $\mu^+\mu^-$ can be produced in the $e^+e^-$ annihilation via two channels:
\[
e^+e^-\Rightarrow \left\{\begin{array}{cc}
\gamma^\ast\\
\psi \\
\end{array}
\right\}\Rightarrow ~\mu^+\mu^-.
\]
The mode via virtual photon $\gamma^\ast$ is the direct electromagnetic production, and another mode is the electromagnetic decay of intermediate on-shell resonance $\psi$. The tree-level Feynman diagram for this process is the coherent summation of the two diagrams in Fig. \ref{treeamplitude}:

\begin{center}
\includegraphics[width=10.5cm]{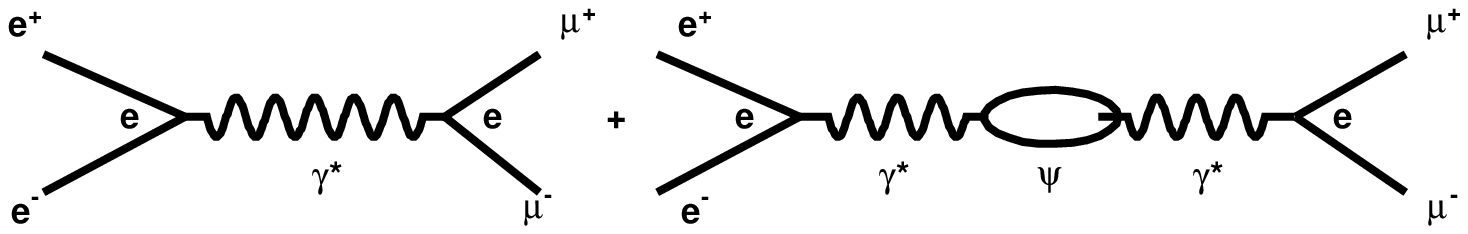}
\figcaption{\label{treeamplitude} Tree-level Feynman diagrams for processes $e^+e^-\to\mu^+\mu^-$ via modes $\gamma^\ast$ (left) and $\psi$ (right). Charge $e$ at the vertex expresses the coupling strength between a lepton and photon.}
\end{center}

Virtual photon propagator $\gamma^\ast$ is unobservable in the experiment, and its role is transferring the electromagnetic interaction between $e^+e^-$ and $\mu^+\mu^-$. Intermediate resonance $\psi$ is a real particle, which is a $c\bar{c}$-bound state with well-defined mass, life-time, spin, and parity $J^{PC}=1^{--}$. Resonances $J/\psi$ and $\psi(3686)$ are identified with the $1S$ and $1P$ levels of the charmonium family predicted by the potential model\cite{highercharmonia}. Unstable $J/\psi$ and $\psi$ will decay into different final states via five modes\cite{kopke}; here, only electromagnetic decay $\psi\to\mu^+\mu^-$ is discussed.

\subsection{Cross-section of $\bm{e^+e^-\to\gamma^\ast\to\mu^+\mu^-}$}

Channel $e^+e^-\to\gamma^\ast\to\mu^+\mu^-$ is a pure QED process, which corresponds to the left diagram in Fig.\ref{treeamplitude}, and the expression of the Born cross-section can be found in any QED text book\cite{peskin}:
\begin{equation}
\sigma_{\gamma^\ast}^0(s)=\frac{4\pi\alpha^2}{3s}.\label{gammatomumu}
\end{equation}

\subsection{Cross-section of $\bm{e^+e^-\to\psi\to\mu^+\mu^-}$}

The channel via intermediate resonance $\psi$ corresponds to the right diagram in Fig.~\ref{treeamplitude}, which concerns the production and decay of $\psi$. This section will provide some description about this mode.

In general, the wavefunction of time for an unstable particle is expressed as a plane wave with a damping amplitude:
\begin{eqnarray}
\Psi(t)&=&\theta(t)\Psi(0)\cdot {\rm e}^{{\rm i}\omega t}\cdot {\rm e}^{-t/2\tau}\nonumber\\
       &=&\theta(t)|\Psi(0)|\cdot {\rm e}^{{\rm i}\delta}\cdot {\rm e}^{-{\rm i}t(M-{\rm i}\varGamma/2)},
\end{eqnarray}
where $\theta(t)$ is a step-function of time, $\Psi(0)$ is the wave function at origin $t=0$, $\omega$ is the circular frequency, $\tau$ is the life-time, and $\delta$ is the intrinsic phase angle of $\Psi(0)$. Here, the relations of mass $M=\omega$ and total decay width $\varGamma=1/\tau$ in natural unit $\hbar=c=1$ are used. For a free particle, its parameters are bare quantities.

Performing the Fourier transformation on $t$ for $\Psi(t)$, the amplitude of an unstable particle is transformed to nonrelativistic wavefunction of energy $W$:
\begin{equation}
{\cal T}(W)=\int_{-\infty}^{+\infty}\Psi(t)\cdot {\rm e}^{{\rm i}Wt}{\rm d}t=\frac{i|\psi(0)|{\rm e}^{{\rm i}\delta}}{(W-M)+i\varGamma/2},
\end{equation}
where the following formula is used:
\begin{equation}
\int_0^{\infty}{\rm e}^{-pt}{\rm d}t=\frac{1}{p},~~~~~~({\rm Re}~ p>0).
\end{equation}
Origin wavefunction $\Psi(0)$ can be determined from the normalization condition and production cross-section\cite{peskin}. Considering a distinct production and decay process with initial state $e^+e^-$ and final state $f$, the corresponding nonrelativistic amplitude is\cite{pdg2014}:
\begin{equation}
{\cal T}_f(W)=\frac{i\sqrt{\varGamma_e\cdot\varGamma_f}{\rm e}^{{\rm i}\delta}}{(W-M)+i\varGamma/2},
\end{equation}
where $\varGamma_e$ and $\varGamma_f$ are the bare electronic and final state widths. For final state $\mu^+\mu^-$, $\varGamma_f=\varGamma_\mu$. The lepton universality implies $\varGamma_e=\varGamma_\mu$ under limit $m_l^2/s\to0$.

The relativistic amplitude can be obtained easily by adopting the physics picture of the Dirac sea. Dirac considered that an antiparticle corresponded to a hole with same mass $M$ but with negative energy state $-W$ in the Dirac sea. Therefore, the relativistic amplitude, which includes particle--antiparticle, is:
\begin{eqnarray}
{\cal T}_f(W)&=&\frac{i\sqrt{\varGamma_e\cdot\varGamma_f}{\rm e}^{{\rm i}\delta}}{(W-M)+i\varGamma/2}+\frac{i\sqrt{\varGamma_e\cdot\varGamma_f}{\rm e}^{{\rm i}\delta}}{(-W-M)+i\varGamma/2}\nonumber\\
&=&\frac{i\sqrt{\varGamma_e\cdot\varGamma_f}(2M-i\varGamma){\rm e}^{{\rm i}\delta}}{W^2-M^2+\varGamma^2/4+i\varGamma M}\nonumber\\
&\approx&\frac{i2M\sqrt{\varGamma_e\varGamma_f}{\rm e}^{{\rm i}\delta}}{W^2-M^2+i\varGamma M}.\label{nonrrlativisticbwam}
\end{eqnarray}
For narrow resonances $J/\psi$ and $\psi(3686)$, the value of $\varGamma$ is assumed much smaller than $M$ and the energy dependence of the total width can be neglected, i.e., $\varGamma$ is treated as a constant.

The Born cross-section for the resonant mode corresponding to the right diagram in Fig.\ref{treeamplitude} is generally written in the Breit--Wigner form:
\begin{equation}
\sigma_{\psi}^0(s)=\frac{4\pi\alpha^2}{3s}|{\cal A}_{\rm BW}|^2,~~~{\cal A}_{\rm BW}=\frac{Fr{\rm e}^{{\rm i}\delta}}{\Delta+ir},\label{psitomumu}
\end{equation}
where the following notations are used:
\begin{eqnarray}
\Delta&=&\frac{s-M^2}{M^2}=t-1,~~~~t=\frac{s}{M^2},\label{deltadefold}\\
r&=&\frac{\varGamma}{M},\\
F&=&\frac{3\sqrt{s\varGamma_e\varGamma_f}}{\alpha\varGamma M}=\frac{3}{\alpha}\sqrt{tB_eB_f}\label{fdefinition}.
\end{eqnarray}
Combination parameter $F$ ensures Eq.~(\ref{psitomumu}) provides the accurate Breit--Wigner cross-section.

Starting with the Van Royen--Weisskopf formula, $\varGamma_e$ can be expressed by the following formula\cite{barbieri,royen,kopke}:
\begin{equation}
\varGamma_e=\frac{16}{3}\pi\alpha^2e_c^2N_c\frac{|R(0)|^2}{M^2}\left(1-\frac{16\alpha_s}{3\pi}\right),\label{gammaemodel}
\end{equation}
where $e_c=2/3$ is the charge of the charm quark in units of electron charge $e$, $N_c=3$ is the number of colors, $\alpha_s$ is the strong coupling constant evaluated at $s=M^2$, and $R(0)$ is the radial wavefunction of $R(t)$ at origin $t=0$. Some phenomenological models can provide a rough estimation for the value of $R(0)$, but its accurate value has to be extracted based on the measurements of $\varGamma_e$ and $\varGamma_f$.

\subsection{Total Born cross-section}

The total production amplitude of $\mu^+\mu^-$ should be a coherent summation of the two channels:
\begin{equation}
{\cal A}_{\rm eff}=1+\frac{Fr{\rm e}^{{\rm i}\delta}}{\Delta+ir}.\label{totalbornamp}
\end{equation}
The total Born cross-section can be written as:
\begin{equation}
\sigma^0(s)=\frac{4\pi\alpha^2}{3s}|{\cal A}_{\rm eff}|^2.\label{totalborncrxt}
\end{equation}

In practical evaluations, the parameter values in the Breit--Wigner cross-section typically adopt the experimental values published in the PDG, which contain the radiative effect\cite{pdg2014,agshamov}. However, the interesting values in physics are the bare ones. The following sections will deduce the total cross-section formula for $e^+e^-\to\gamma^\ast/\psi\to\mu^+\mu^-$, in which all the parameters are bare quantities. Based on this formula, the bare parameter values can be extracted by fitting the measured cross-section.

\section{Vacuum polarization correction}\label{vpsection}

From the viewpoint of quantum field theory, two charged particles interact by exchanging quanta of the electro-magnetic field, which corresponds to the virtual photon propagator between the two charges. The VP effect modifies the photon propagator, which is equivalent to a change in the coupling strength between two charges. In the one-particle-irreducible (1PI) chain approximation, an infinite series of 1PI diagrams is summed, and the photon propagator is modified by the VP correction in following manner \cite{peskin}:
\begin{equation}
\gamma^\ast:~~\frac{-ig_{\mu\nu}}{q^2}~~\longrightarrow~~\tilde{\gamma}^\ast:~~\frac{-ig_{\mu\nu}}{q^2[1-\Pi(q^2)]},\label{phtontophotonvpform}
\end{equation}
where $g_{\mu\nu}$ is the metric tensor and $\Pi(q^2)$ is the VP function. For the $e^+e^-$ annihilation process, $q^2=s$. Eq.~(\ref{phtontophotonvpform}) can be expressed graphically as the bare propagator, $\gamma^\ast$, is modified to be the full propagator, $\tilde{\gamma}^\ast$:

\begin{center}
\includegraphics[width=10cm]{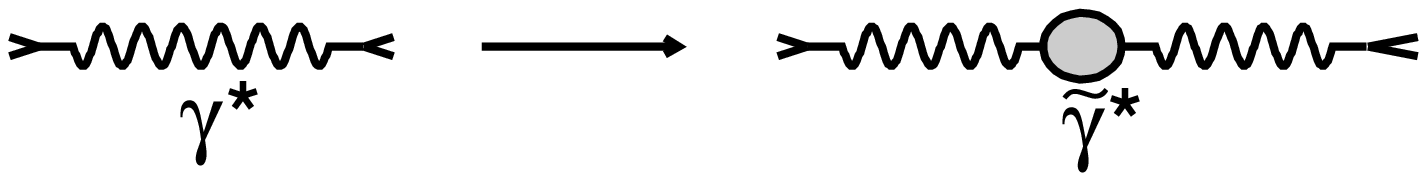}
\figcaption{\label{phtontophotonvp} Bare propagator $\gamma^\ast$ is replaced by full propagator $\tilde{\gamma}^\ast$ with the VP correction.}
\end{center}

The original algorithm of $\Pi(s)$ is an infinite integral of fermion-loops (leptons and quarks) in the four-momentum space. The integral for the QED lepton-loops ($e^+e^-$, $\mu^+\mu^-$, $\tau^+\tau^-$) can be calculated perturbatively according to the Feynman rules\cite{peskin,greiner}. The divergence of the infinite integral is canceled by electric charge renormalization $e_0\to\sqrt{Z_3}e_0=e$, where $e_0$ is the bare electric charge in the original Lagrangian, $e$ is the physical charge, and the renormalization constant is
\begin{equation}
Z_3\equiv\frac{1}{1-\Pi(0)},~~~(\Pi(0)\to\infty).
\end{equation}
The remaining finite part of $\Pi(s)$ is $\hat\Pi(s)=\Pi(s)-\Pi(0)$, which is used to define running coupling constant $\alpha(s)$ to the leading order:
\begin{equation}
\alpha(s)=\frac{e_0^2/4\pi}{1-\Pi(s)}=\frac{\alpha}{1-[\Pi(s)-\Pi(0)]}\equiv\frac{\alpha}{1-\hat\Pi(s)}.\label{runingalpha}
\end{equation}
This formula expresses an important physics characteristic: finite part $\hat\Pi(s)$ in Eq.~(\ref{runingalpha}) is not the entire VP function; infinite part $\Pi(0)$ is absorbed into the definition of physical charge $e$.

After the charge renormalization, the effect of the VP correction can be explained as bare charge $e_0$ is redefined as physical charge $e$ and simultaneously fine-structure constant $\alpha$ is replaced by effective energy-dependent running coupling factor $\alpha(s)$. Therefore, finite part $1-\hat\Pi(s)$ of the VP factor should be combined with $\alpha$ to yield effective running constant $\alpha(s)$. Thus, $\alpha$ and $1-\hat\Pi(s)$ should not be separated in the physical explanations and practical calculations.

In one-photon exchange and chain approximation, the finite part of VP function $\hat\Pi(s)$ can be expressed as the summation of all of fermion-loop contributions\cite{slac4160,slac5160,agshamov}:
\begin{equation}
\hat\Pi(s)=\sum [\Pi_{l\bar{l}}(s)+\Pi_{q\bar{q}}(s)],
\end{equation}
where $l\bar{l}=e^-e^+,\mu^-\mu^+,\tau^-\tau^+$, and $q\bar{q}=u\bar{u}$, $d\bar{d}$, $s\bar{s}$, $c\bar{c}$, $b\bar{b}$, $t\bar{t}$. The QED terms of the lepton-loops can be calculated analytically\cite{peskin,greiner}. However, for the QCD quark-loops, analytic calculations cannot be used owing to the strong nonperturbative interaction. The solution for this issue is to use the optical theorem and dispersion relation\cite{ncabibbo,ecremmer}.

\begin{figure*}[!t]
\centering
\includegraphics[width=6.2cm]{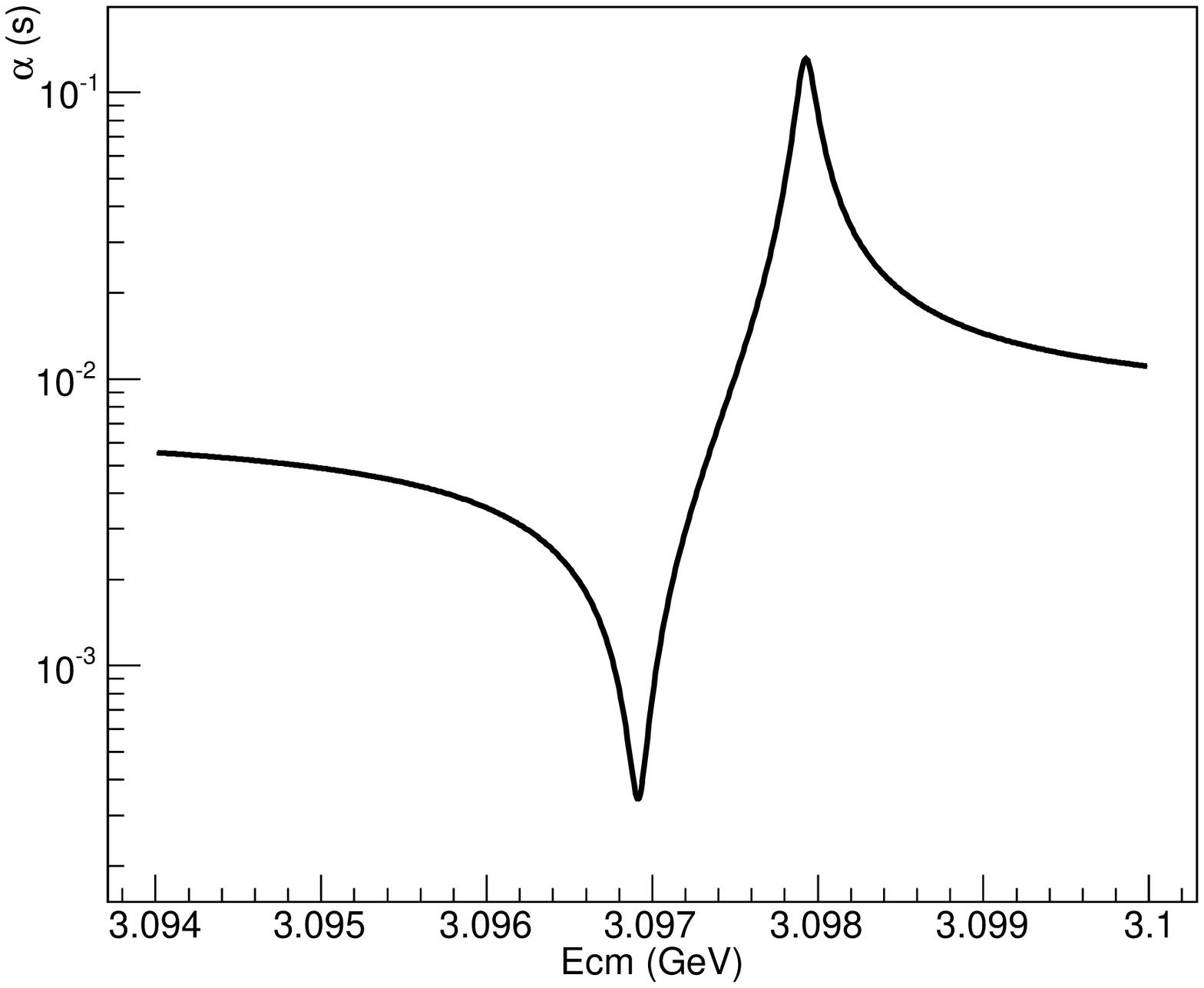}
\includegraphics[width=6.2cm]{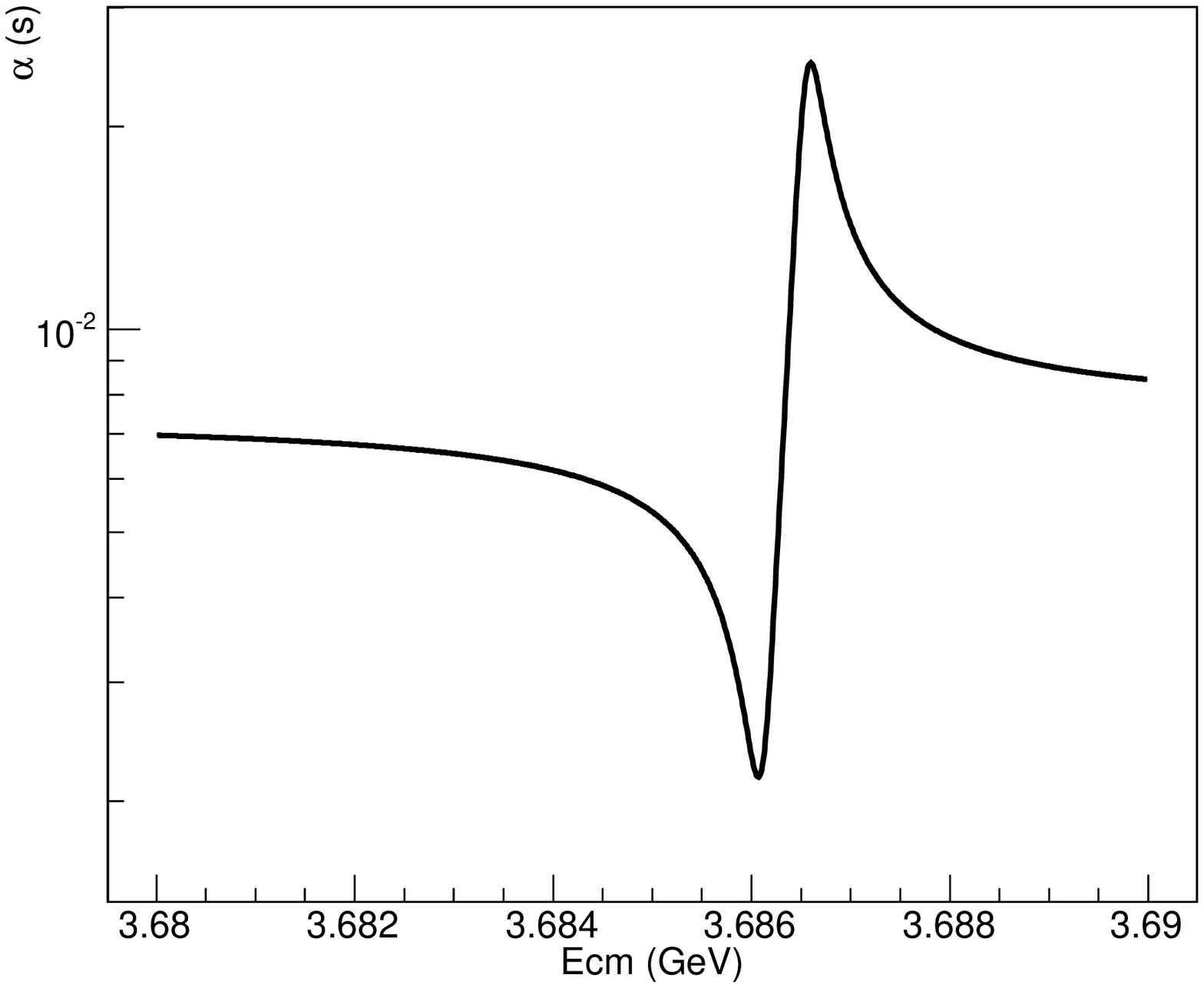}
\figcaption{\label{alphas} Energy dependence of $\alpha(s)$ around $J/\psi$ (left) and $\psi(3686)$ (right).}
\end{figure*}

The optical theorem relates the imaginary part of the QCD component of the photon self-energy to the inclusive hadronic Born cross-section\cite{ecremmer}:
\begin{equation}
{\rm Im}\Pi_{q\bar{q}}(s)=\frac{s}{4\pi\alpha}\sigma_{\rm had}^0(s).\label{opticaltheome}
\end{equation}
The dispersion relation relates the QCD contribution of the VP function to the integral of the imaginary part of the VP function about the quark-loops:
\begin{equation}
\Pi_{q\bar{q}}(s)=\frac{s}{\pi}\int_0^\infty\frac{{\rm Im}\Pi_{q\bar{q}}(s')}{s'(s'-s-i\epsilon)}{\rm d}s'.\label{dispersion}
\end{equation}
Inserting Eq.~(\ref{opticaltheome}) in Eq.~(\ref{dispersion}), the nonperturbative QCD VP term can be calculated using the hadronic cross-section,
\begin{equation}
\Pi_{q\bar{q}}(s)=\frac{s}{4\pi^2\alpha}\int_0^\infty\frac{\sigma_{\rm had}^0(s')}{s'-s-i\epsilon}{\rm d}s'.
\end{equation}
If the interference between the inclusive continuum and resonant hadronic states can be neglected, the contribution of the quark-loops can be written as:
\begin{equation}
\Pi_{q\bar{q}}(s)=\Pi_{\rm con}(s)+\Pi_{\rm res}(s).\label{vpconres}
\end{equation}
$\Pi_{\rm con}(s)$ can be calculated by the numerical integral:
\begin{equation}
\Pi_{\rm con}(s)=\frac{\alpha}{3\pi}\int_0^\infty\frac{\tilde{R}(s')}{s'-s-i\epsilon}{\rm d}s'.
\end{equation}
Generally, $\tilde{R}(s)$ uses experimental values below 5 GeV\cite{kedrrlow,kedrrhigh,besiir2004}, whereas $\tilde{R}(s)$ adopts the perturbative QCD (pQCD) prediction above 5 GeV.

$\Pi_{\rm res}(s)$ includes all the contributions of the resonances with $J^{PC}=1^{--}$. If the interference between different resonances having the same decay final states are neglected for simplicity, resonant cross-section $\sigma_{\rm res}^0(s)$ can be written as the summation of the Breit--Wigner cross-sections:
\begin{equation}
\sigma_{\rm res}^0(s)=\sum_j\frac{12\pi\varGamma_{ej}\varGamma_j}{(s-M_j^2)^2+M_j^2\varGamma_j^2},~~~(j=\rho,\omega...\psi...),\label{bwcrxt}
\end{equation}
and the final analytical result is:
\begin{eqnarray}
\Pi_{\rm res}(s)&=&\frac{s}{4\pi^2\alpha}\int_0^\infty\frac{\sigma_{\rm res}^0(s')}{s'-s-i\epsilon}{\rm d}s'\nonumber\\
&=&\sum_j\frac{3s}{\alpha}\frac{\varGamma_{ej}}{M_j}\frac{1}{s-M_j^2+iM_j\varGamma_j}.\label{pires}
\end{eqnarray}
In the vicinity of $J/\psi$ and $\psi(3686)$, their overlap can be neglected and only one resonance needs to be considered. However, in higher charmonia regions, wide $\psi(4040)$, $\psi(4160)$ and $\psi(4415)$ overlap significantly, and all their contributions and interference effects should be included\cite{plb6602008315}.

Figure~\ref{alphas} exhibits the energy dependence of running coupling constant $\alpha(s)$ expressed by Eq.~(\ref{runingalpha}) around resonances $J/\psi$ and $\psi(3686)$. The resonant shape of $\alpha(s)$ is due to the virtual VP effect, instead of the real resonance produced.

It should be noticed that in experiment measurements, there is no strict partition between the continuum and resonant states, as expressed in Eq.~(\ref{sigmaconres}). For example, observed final state $\pi^+\pi^-$ may be direct production $e^+e^-\to\pi^+\pi^-$ or via intermediate mode $e^+e^-\to\rho^0\to\pi^+\pi^-$. Therefore, Eqs.~(\ref{sigmaconres}) and ~(\ref{vpconres}) are only roughly divided for simplicity.

It should be stressed that the dispersion relation and optical theorem merely provide a practical algorithm for calculating QCD nonperturbative VP function $\Pi_{q\bar{q}}(s)$, which does not provide extra physics explanation. However, the procedures for calculating $\Pi_{q\bar{q}}(s)$ from the dispersion relation and optical theorem may be misleading. Some users considered that cross-sections $\sigma_{\rm con}^0(s)$ and $\sigma_{\rm res}^0(s)$ in the expressions of $\Pi_{q\bar{q}}(s)$ imply that the VP effect also produces real continuum and resonant hadronic states in the virtual photon propagator. In fact, the fermion-loop integral of the VP function is the virtual quantum fluctuation by its definition, and it does not have characteristic quantum numbers (such as, mass, spin, parities), which are necessary for any real particle. A real physics state must be able to be measured in detectors, but the fermion-loops with infinite four-momentum fluctuations in the VP cannot be observed.

In general, the Born cross-sections of the $\gamma^\ast$ mode and intermediate $\psi$ mode are proportional to $\alpha^2$. Considering the VP effect, running coupling constant $\alpha(s)$ leads to an additional energy-dependence of the cross-section. Moreover, for the energy region around $J/\psi$ and $\psi(3686)$, the value of $\Pi_{\rm res}(s)$ is very sensitive to $s$, $\varGamma_e$, and $\varGamma$, which implies that the bare values of $\varGamma_e$ and $\varGamma$ will influence the line-shape of $e^+e^-\to\gamma^\ast/\psi\to\mu^+\mu^-$ significantly.

\section{Effective leptonic width \label{effeewidth}}

\begin{figure*}[!bp]
\centering
\includegraphics[width=6.2cm]{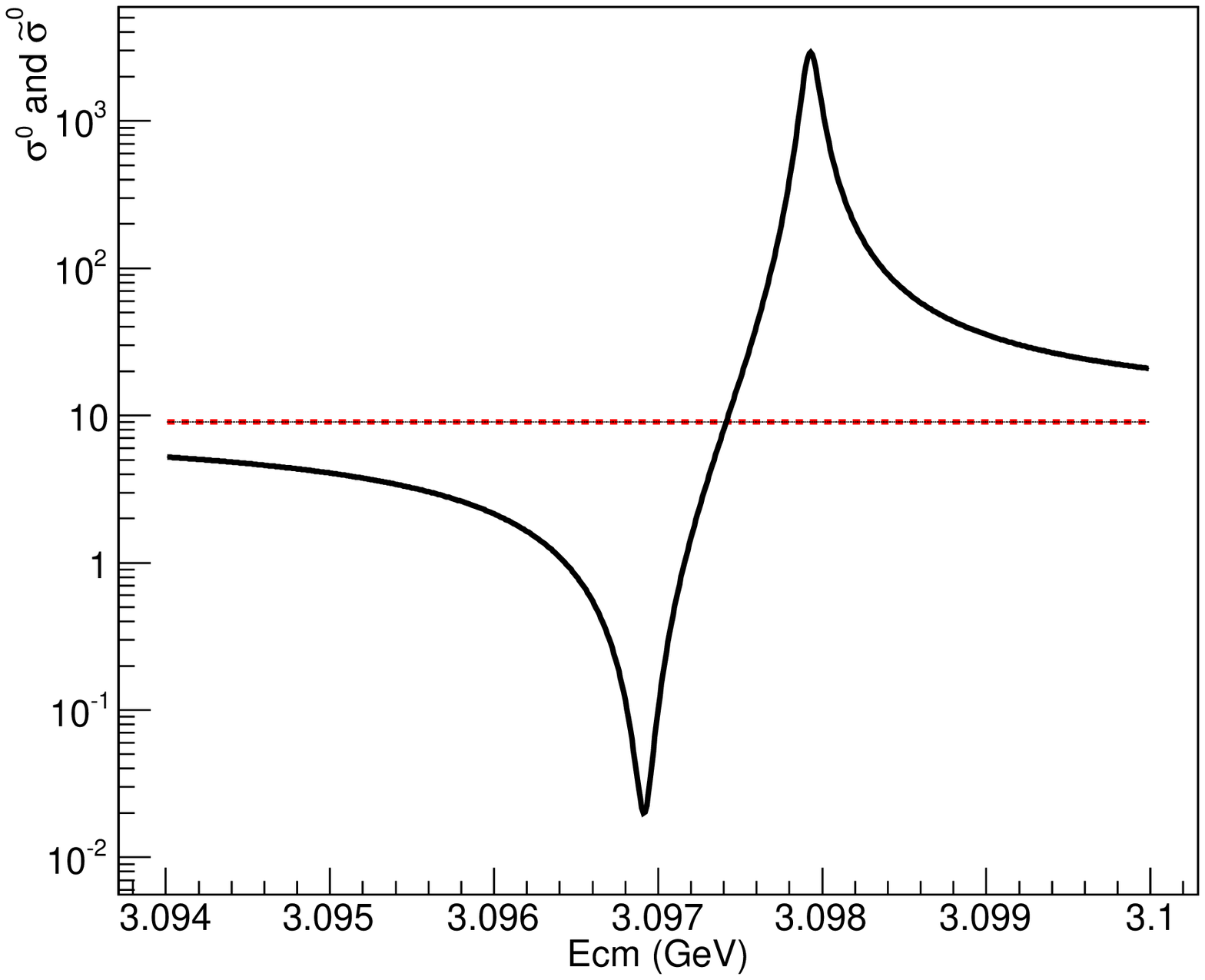}
\includegraphics[width=6.2cm]{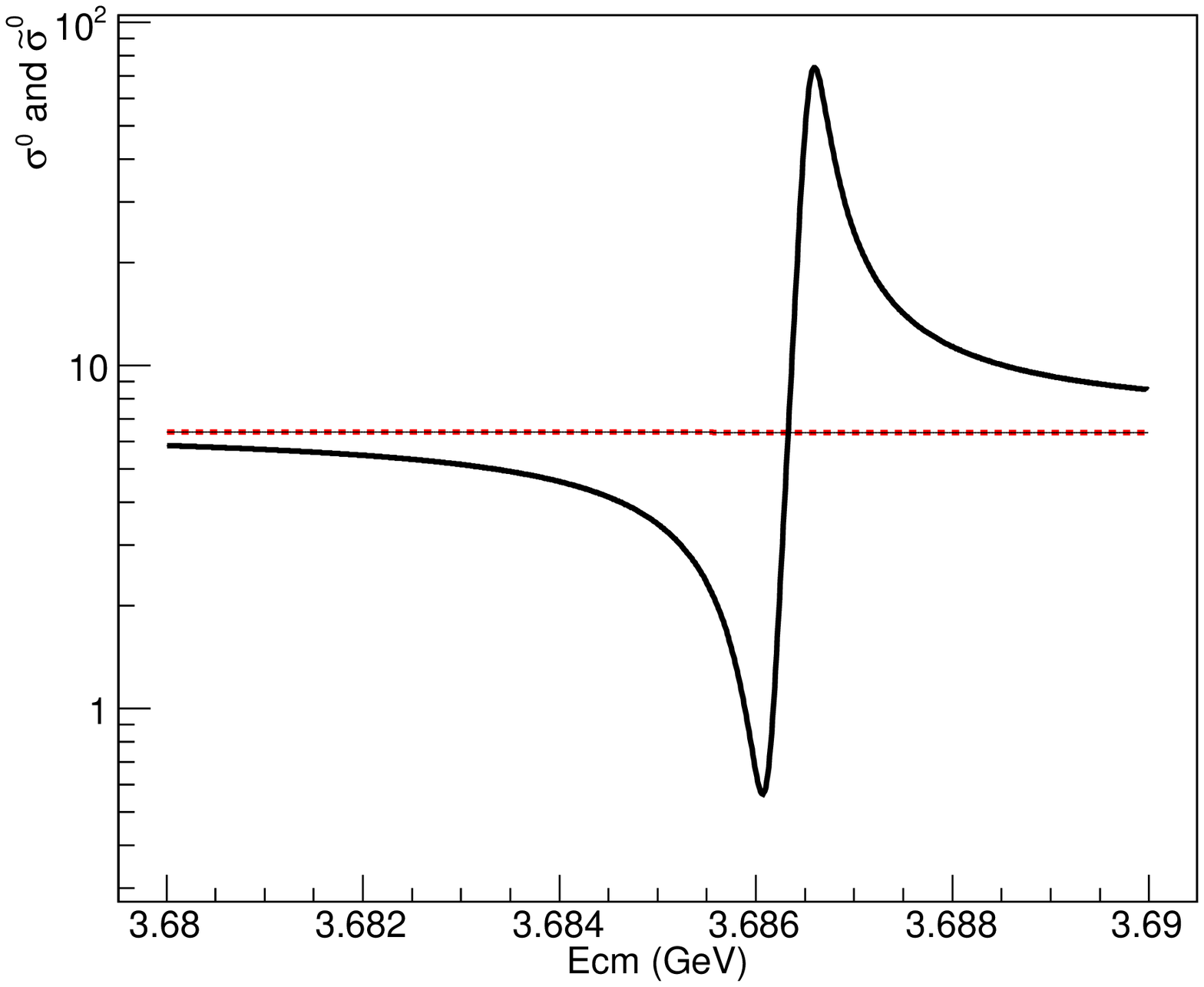}
\figcaption{\label{eemumugammachannel}(color online) Line-shape of $\sigma_{\gamma\ast}^0(s)$ (dashed line) and $\tilde{\sigma}_{\gamma^\ast}^0(s)$ (solid line) in the vicinity of $J/\psi$ (left) and $\psi(3686)$ (right).}
\end{figure*}

In most references, the value of the electron width in the Breit--Wigner cross section adopts experimental partial width $\varGamma_e^{ex}$ (which is represented as $\varGamma_e$ in the PDG without declaring), with the VP effect being absorbed into the electron width. There are two different conventions for $\varGamma_e^{ex}$.

In reference \cite{ystsai}, the experimental electron width is defined as:
\begin{equation}\label{gammaetsai}
\varGamma_e^{ex}=\frac{\varGamma_e}{|1-\hat\Pi(M^2)|^2},
\end{equation}
where the entire VP function is absorbed in $\varGamma_e^{ex}$. In reference \cite{agshamov}, the following definition is adopted:
\begin{equation}\label{gammaekedr}
\varGamma_e^{ex}=\frac{\varGamma_e}{|1-\hat\Pi_0(M^2)|^2},~~\hat\Pi_0(s)=\hat\Pi_{\rm QED}(s)+\hat\Pi_{\rm QCD}(s).
\end{equation}
This convention implies that electron width $\varGamma_e^{ex}$ absorbs the contribution of the non-resonant components only, whereas the resonant component of the VP correction is absorbed in parameters $M$ and $\varGamma$, introducing dressed values $\tilde{M}$ and $\tilde{\varGamma}$. Thus, $\tilde{M}$ and $\tilde{\varGamma}$ deviate the original physical relevance of the mass and total width (life-time). The conventions in Eqs.~(\ref{gammaetsai}) and ~(\ref{gammaekedr}) are not equivalent for $\varGamma_e^{ex}$. It is important to clarify which convention is adopted for the appropriate application of $\varGamma_e^{ex}$ cited in the PDG.

It is seen from the discussion in the above section, it is not necessary to introduce quantity $\varGamma_e^{ex}$ in the expression of the cross-section if $\alpha$ is replaced by $\alpha(s)$. The following sections will discuss this in more detail. Using $\alpha(s)$ to replace $\alpha$ can keep the bare value $\varGamma_e$ in the analysis, which is more natural for understanding the VP effect than introducing $\varGamma_e^{ex}$. However, if bare value $\varGamma_e$ is measured using the scheme proposed in this paper, one may obtain $\varGamma_e^{ex}$ by the definition in Eq.~(\ref{gammaetsai}) or Eq.~(\ref{gammaekedr}) and extract radial wave function $R(0)$ according to Eq.~(\ref{gammaemodel}).

\section{VP-modified Born cross-section}

\begin{figure*}[t]
\centering
\includegraphics[width=6.2cm]{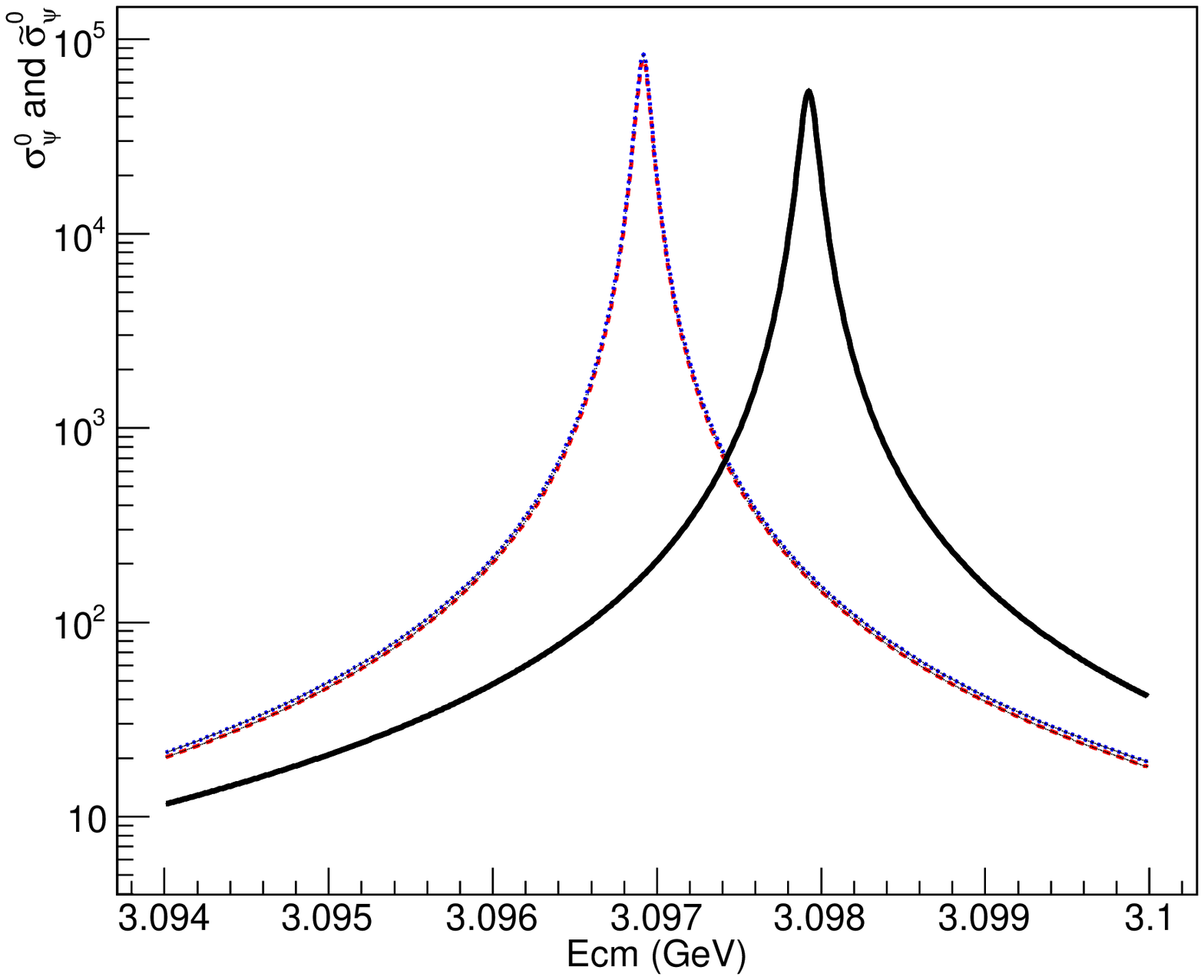}
\includegraphics[width=6.2cm]{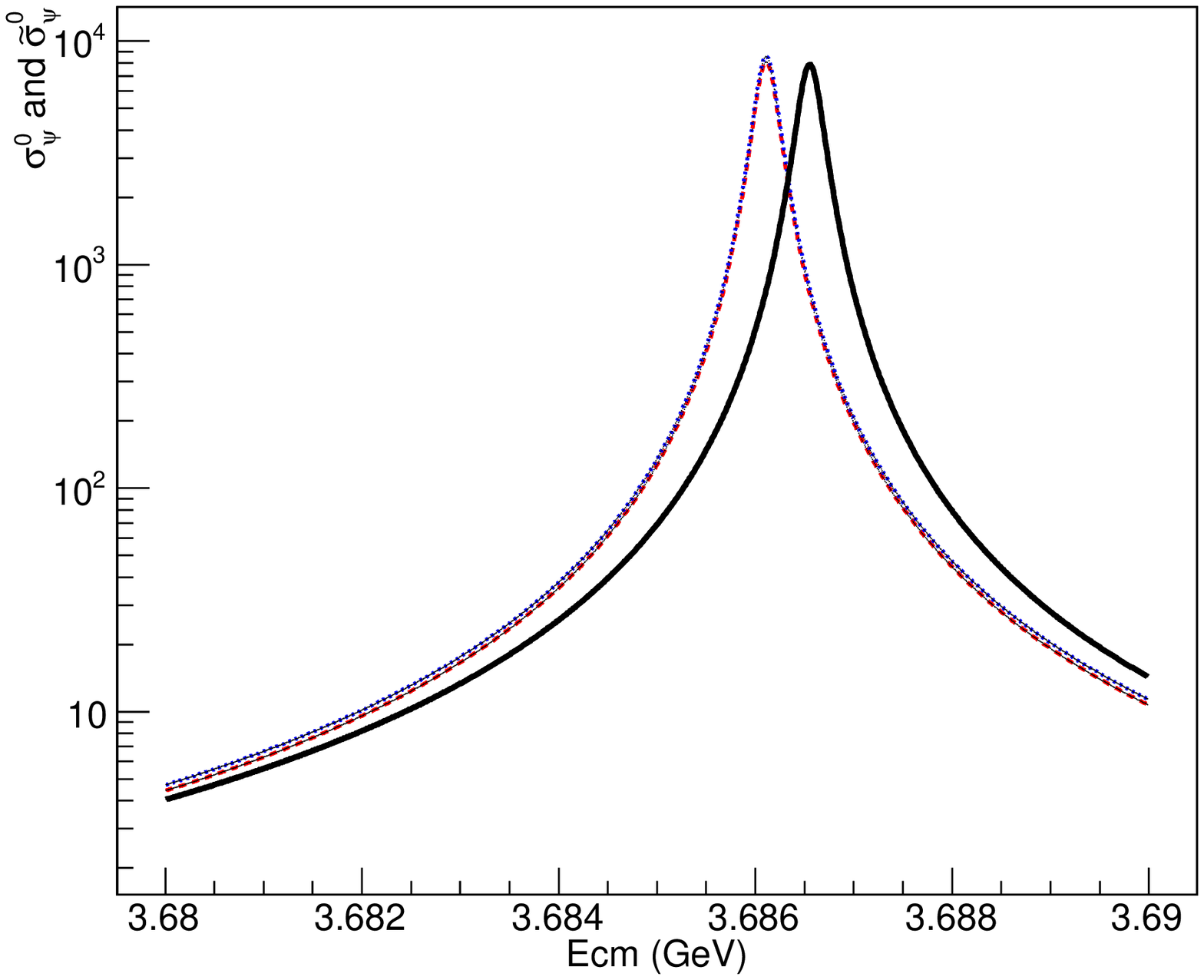}
\figcaption{\label{vpmodifidbw}(color online) Line-shape comparison of resonant channels $e^+e^-\to J/\psi\to\mu^+\mu^-$ (left) and $e^+e^-\to \psi(3686)\to\mu^+\mu^-$ (right) between Born-level Breit--Wigner cross-section $\sigma_{\psi}^0(s)$ (dashed line) and VP-modified cross-section $\tilde{\sigma}_{\psi}^0(s)$ (solid line).}
\end{figure*}

From the viewpoint of Feynman diagrams, the VP correction modifies the photon propagator, which can be understood from another perspective: the VP effect modifies fine structure constant $\alpha$ to running coupling constant $\alpha(s)$. In this section, the single and double VP effects will be discussed and their differences will be compared numerically.

The VP-corrected total Born cross-section is:
\begin{equation}
\tilde{\sigma}^0(s)=\frac{\sigma^0(s)}{|1-\hat\Pi(s)|^2}=\frac{\sigma_{\gamma^\ast}^0(s)+\sigma_{\psi}^0(s)}{|1-\hat\Pi(s)|^2}.
\end{equation}
The next two sections will discuss the effect of VP on $\sigma_{\gamma^\ast}^0(s)$ and $\sigma_{\psi}^0(s)$, respectively.

\subsection{VP-modified cross-section of $\bm{\gamma^\ast}$ channel}

Born cross-section $\sigma_{\gamma\ast}^0(s)$ of $\gamma^\ast$ channel expressed in Eq.~(\ref{gammatomumu}) is a smooth function of the energy. When the VP correction is applied to it,
\begin{equation}
\sigma_{\gamma^\ast}^0(s)~\to~\tilde{\sigma}_{\gamma^\ast}^0(s)=\frac{\sigma_{\gamma\ast}^0(s)}{|1-\hat\Pi(s)|^2}=\frac{4\pi\alpha^2(s)}{3s}.\label{convp}
\end{equation}

Figure~\ref{eemumugammachannel} shows the line-shapes of $\sigma_{\gamma\ast}^0(s)$ given in Eq.~(\ref{gammatomumu}) and of $\tilde{\sigma}_{\gamma^\ast}^0(s)$ in Eq.~(\ref{convp}). The line-shape of $\sigma_{\gamma\ast}^0(s)$ is smooth for $s$, and $\tilde{\sigma}_{\gamma^\ast}^0(s)$ gives the obvious resonant structure. Clearly, the resonant structure of $\tilde{\sigma}_{\gamma^\ast}^0(s)$ is owing to the VP effect or the sensitive energy-dependence of $\alpha(s)$ in the vicinity of $\psi$, and $\tilde{\sigma}_{\gamma^\ast}^0(s) < \sigma_{\gamma\ast}^0(s)$ for $s<M^2$, $\tilde{\sigma}_{\gamma^\ast}^0(s) > \sigma_{\gamma\ast}^0(s)$ for $s>M^2$. Thus, the resonant shape of the $\gamma^\ast$ channel cross-section does not imply that real resonant state $J/\psi$ or $\psi(3686)$ is produced but that resonant component $\Pi_{\rm res}(s)$ affects the VP function. In the vicinities of narrow resonances, both Born cross-section $\sigma^0(s)$ expressed in Eq.~(\ref{totalborncrxt}) and VP function $\hat\Pi(s)$ are sensitive to energy. Therefore, the energy dependence of effective cross-section $\tilde{\sigma}^0(s)$ is not only determined by $\sigma^0(s)$ but also by $\Pi_{\rm res}(s)$ or $\alpha(s)$.

\subsection{VP-modified cross-section of $\psi$ channel}

Generally, the cross-section of a resonance is expressed in the Breit--Wigner form. If the value of the electron width adopts bare value $\varGamma_e$, the effective Breit--Wigner cross-section is modified by the VP correction. The reference \cite{ystsai} adopted the convention defined by Eq.~(\ref{gammaetsai}), which corresponds to the VP effect-modified Breit--Wigner cross-section:
\begin{equation}
\tilde{\sigma}_{\psi}^0(s)=\frac{\sigma_{\psi}^0(s)}{|1-\hat\Pi(M^2)|^2}.\label{bwvpold}
\end{equation}
The numerator and denominator in Eq.~(\ref{bwvpold}) are evaluated at different energy scales; the numerator is evaluated at $s$, and the denominator is evaluated at peak $M^2$. It is inappropriate to make line-shape scan measurements in the vicinity of $J/\psi$ and $\psi(3686)$ because most energy points $s_i$ deviate from peak value $M^2$. In fact, a more natural VP correction for Breit--Wigner cross-section $\sigma_{\psi}^0(s)$ should be
\begin{equation}
\tilde{\sigma}_{\psi}^0(s)=\frac{\sigma_{\psi}^0(s)}{|1-\hat\Pi(s)|^2},\label{bwvpnew}
\end{equation}
which corresponds to the convention:
\begin{equation}
\varGamma_e^{ex}(s)=\frac{\varGamma_e}{|1-\hat\Pi(s)|^2},\label{huconvention}
\end{equation}
which according to Eq.~(\ref{gammaemodel}) and Eq.~(\ref{runingalpha} requires VP effect-modified $\varGamma_e$ to be energy-dependent:
\begin{equation}
\varGamma_e\to\tilde{\varGamma}_e(s)=\frac{16}{3}\pi[\alpha(s)]^2e_c^2N_c\frac{|R(0)|^2}{M^2}
\left(1-\frac{16\alpha_s}{3\pi}\right).\label{gammaemodelalphas}
\end{equation}

Figure~\ref{vpmodifidbw} shows the line-shape comparison of $\sigma_{\psi}^0(s)$ defined in Eq.~(\ref{psitomumu}) and $\tilde{\sigma}_{\psi}^0(s)$ defined in Eq.~(\ref{bwvpold}) and Eq.~(\ref{bwvpnew}) for $J/\psi$ and $\psi(3686)$, respectively. In the calculations for Fig.\ref{vpmodifidbw}, $M$ and $\varGamma$ adopt the PDG values, whereas $\varGamma_e$ uses theoretical values $\varGamma_e=4.8$ keV for $J/\psi$ and $\varGamma_e=2.1$ keV for $\psi(3686)$\cite{eichten}. The difference in the line-shapes based on Eqs.~(\ref{psitomumu}) and ~(\ref{bwvpold}) is small. The peak positions of $\sigma_{\psi}^0(s)$ and $\tilde{\sigma}_{\psi}^0(s)$ defined by Eq.~(\ref{bwvpold}) are the same, and the relative difference in their cross-sections at the peak is approximately $6\%$ for both $J/\psi$ and $\psi(3686)$. The shift in the peak positions between $\sigma_{\psi}^0(s)$ and $\tilde{\sigma}_{\psi}^0(s)$ defined by Eq.~(\ref{bwvpnew}) is approximately $1.0$ Mev and $0.4$ MeV, and the relative difference in their cross-section at the peak is approximately $31\%$ and $3\%$ for $J/\psi$ and $\psi(3686)$, respectively. $J/\psi$ is narrower than $\psi(3686)$, and thus, the shift in the vicinity of $J/\psi$ is much larger than that near $\psi(3686)$. The line-shape of the VP-modified Breit--Wigner cross-section adopting Eq.~(\ref{bwvpold}) and Eq.~(\ref{bwvpnew}) is different. It is clear that adopting Eq.~(\ref{bwvpnew}) is reasonable, and it is consistent with the VP correction to the $\gamma^\ast$ channel, see Eq.~(\ref{convp}).

\subsection{Single VP correction case}

The Feynman diagram with a single VP correction is shown in Fig.~\ref{singlevpamplitude}, where $e$ at the vertex is the electron charge, which represents the coupling strength between the leptons ($e^\pm$ or $\mu^\pm$) and photon ($\gamma^\ast$). The grey bubble represents the VP correction in the 1PI approximation, and the hollow oval represents resonance $\psi$. For the $\psi$ channel in the Feynman diagram in Fig.~\ref{singlevpamplitude}, only the virtual photon propagator between the initial $e^+e^-$ and intermediary $\psi$ is corrected by the VP. There is no VP correction for the virtual photon between $\psi$ and final state $\mu^+\mu^-$, which is same as the traditional treatment, i.e., only a single VP correction is considered for the $\psi$ channel.

\begin{center}
\includegraphics[width=9cm]{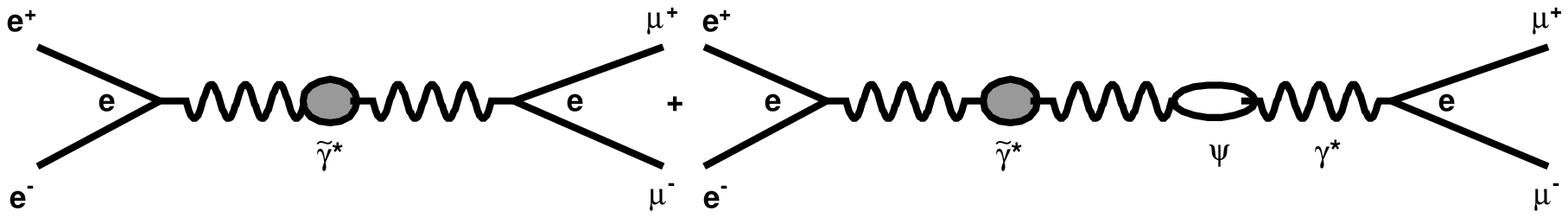}
\figcaption{\label{singlevpamplitude} Feynman diagram with a single VP correction.}
\end{center}

\begin{figure*}[t]
\centering
\includegraphics[width=6.2cm]{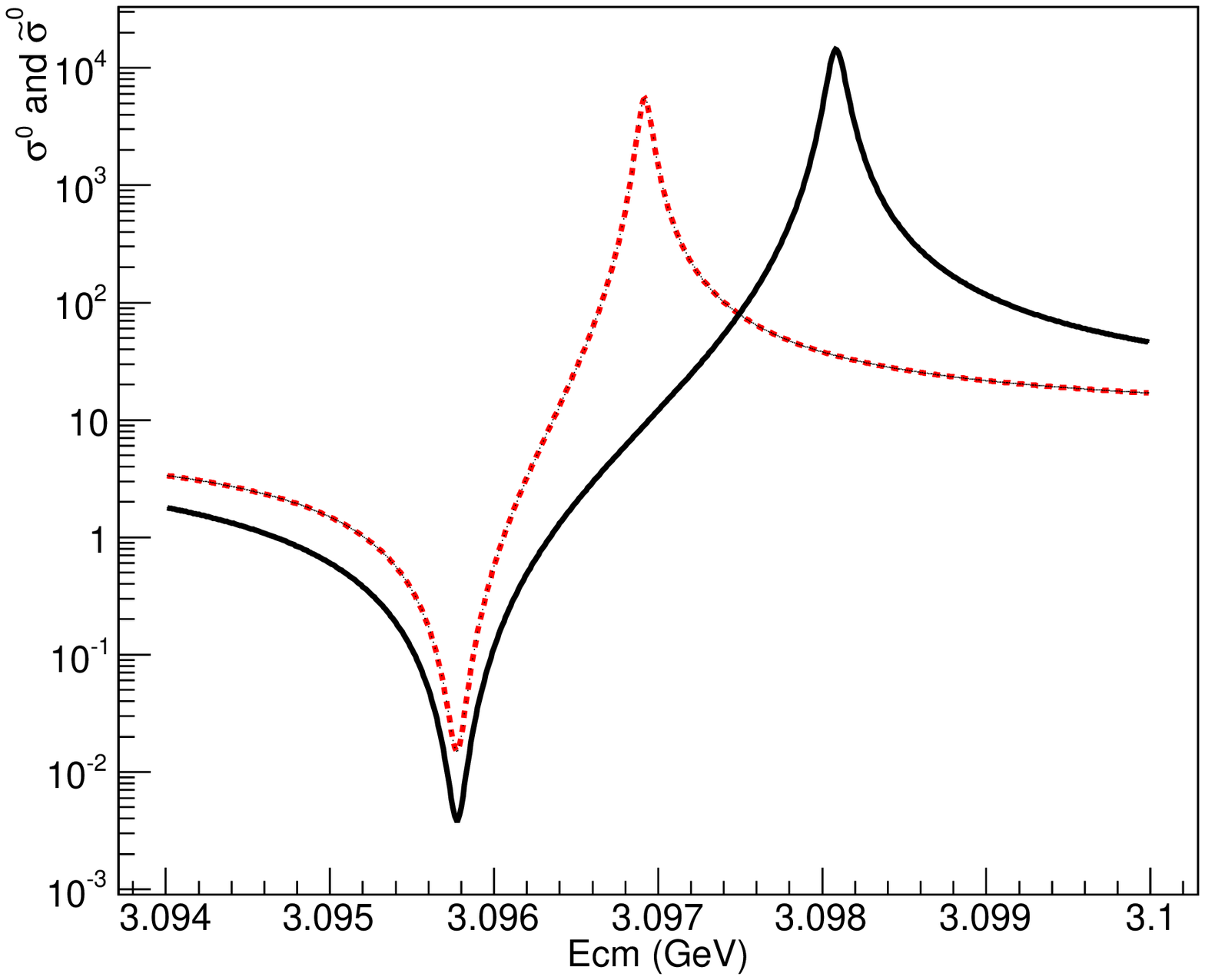}
\includegraphics[width=6.2cm]{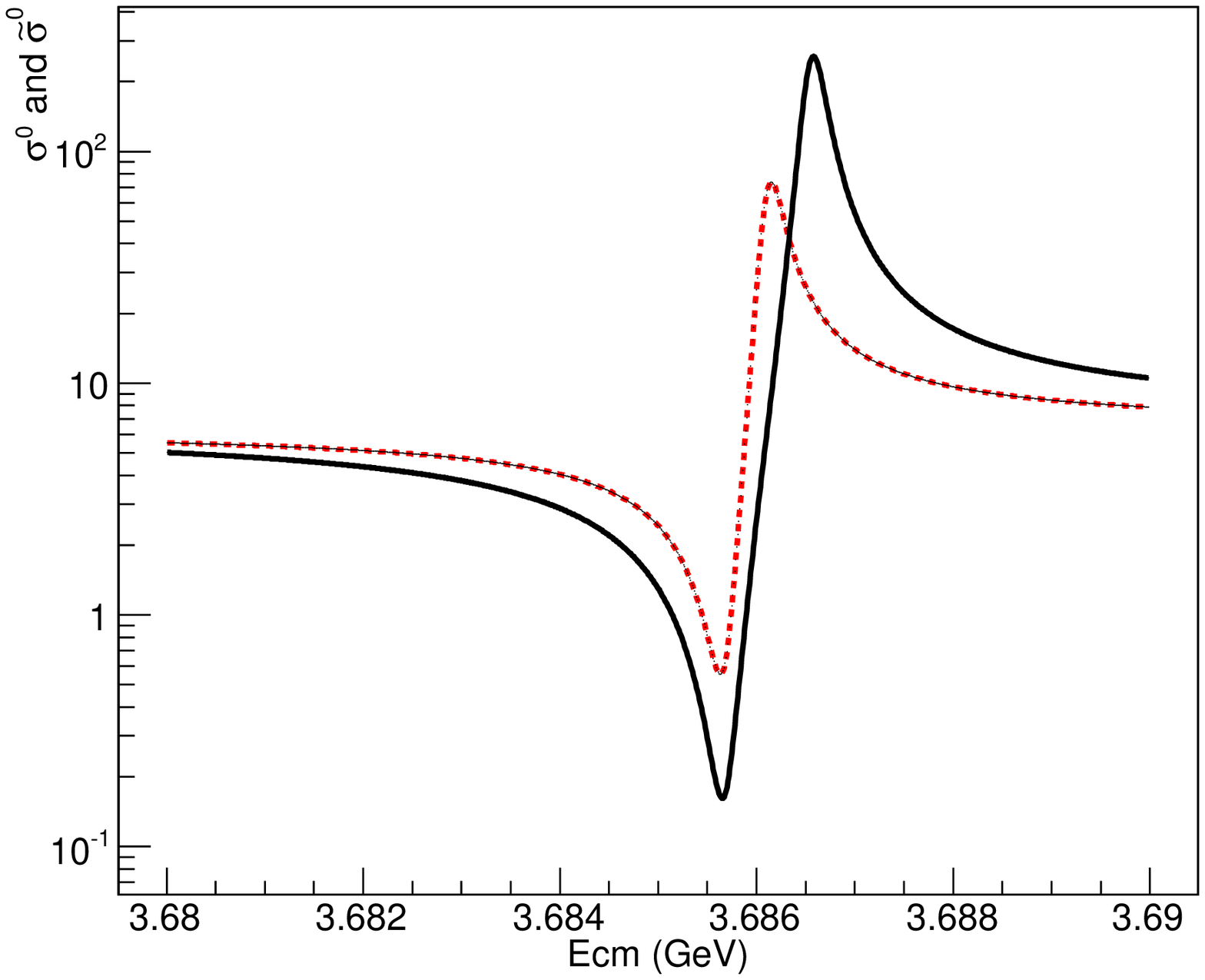}
\figcaption{\label{eemmborncrxtsgvp} Line-shape of $\sigma^0(s)$ by Eq.~(\ref{totalborncrxt}) (dashed line) and $\tilde{\sigma}^0(s)$ expressed by Eq.~(\ref{origbornvpcor}) (solid line) in the vicinity of $J/\psi$ (left) and $\psi(3686)$ (right).}
\end{figure*}

A coherent amplitude is given by sum of two diagrams:
\begin{equation}
\tilde{{\cal A}}_{\rm eff}\sim\frac{1}{1-\hat\Pi(s)}(1+\frac{Fr{\rm e}^{{\rm i}\delta}}{\Delta+ir}).\label{singlevpamplitudeformula}
\end{equation}
Considering the VP effect and that the electromagnetic coupling strength still expresses as $\alpha$, the Born cross-section is modified as the following expression:
\begin{equation}\label{origbornvpcor}
\sigma^0(s)\rightarrow\tilde{\sigma}^0(s)=\frac{4\pi\alpha^2}{3s}|\tilde{{\cal A}}_{\rm eff}|^2=\frac{\sigma^0(s)}{|1-\hat\Pi(s)|^2},
\end{equation}
where $\sigma^0(s)$ is given by Eq.~(\ref{totalborncrxt}). The energy dependence of $\sigma^0(s)$ and $\tilde{\sigma}^0(s)$ in the vicinity of $J/\psi$ and $\psi(3686)$ is displayed in Fig.~\ref{eemmborncrxtsgvp}. It is clear that the VP correction or equivalent $\alpha(s)$ distorts the line-shape of the original resonant structure of $\sigma^0(s)$.

The Feynman diagram with a single VP correction in Fig.~\ref{singlevpamplitude} can also be replotted as Fig.~\ref{singlelevpamplitude} equivalently, which has the same topological structure
as the tree level in Fig.~\ref{treeamplitude}. The black-dot at the vertex is effective running electron charge:
\begin{equation}
e^2(s)=\frac{e^2}{|1-\hat\Pi(s)|}.\label{runninge}
\end{equation}

\begin{center}
\includegraphics[width=10.5cm]{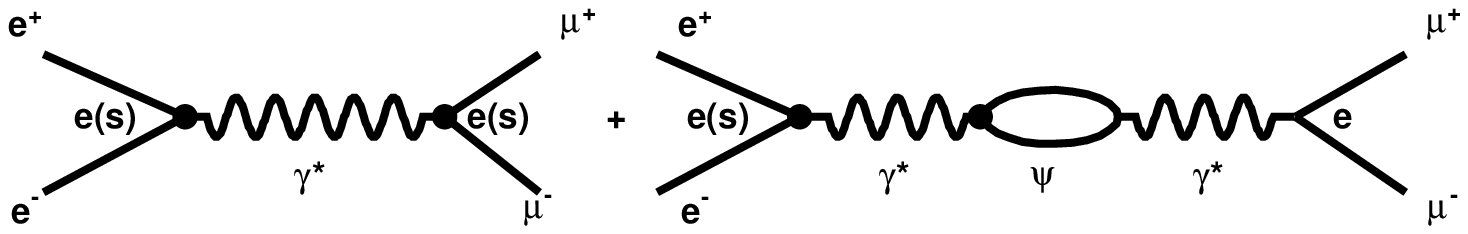}
\figcaption{\label{singlelevpamplitude} Equivalent Feynman diagram of Fig.~\ref{singlevpamplitude}. The single VP correction is absorbed into effective electric charge $e(s)$ defined in Eq.~(\ref{runninge}).}
\end{center}

For the right Feynman diagram of channel $e^+e^-\to\psi\to\mu^+\mu^-$ in Fig.~\ref{singlevpamplitude} or Fig.~\ref{singlelevpamplitude}, coupling strength of three-line vertex $e^+e^-\gamma^{\ast}$ is $e(s)$ corresponding to $\alpha(s)$, and for $\mu^+\mu^-\gamma^{\ast}$, it is $e$ corresponding to $\alpha$:
\begin{equation}
\alpha=\frac{e^2}{4\pi},~~~~{\rm and}~~~~\alpha(s)=\frac{e^2(s)}{4\pi}.
\end{equation}

\subsection{Double VP correction case}

In the quantum field theory, processes $e^+e^-\to\mu^+\mu^-$ and $\mu^+\mu^-\to e^+e^-$ should be invariant under time reversal $T\rightleftarrows-T$, and both processes have the same cross-section if masses $m_e$ and $m_\mu$ can be neglected compared to energy $\sqrt{s}$. However, the right Feynman diagrams in Fig.~\ref{singlevpamplitude} and Fig.~\ref{singlelevpamplitude} violate this basic requirement. This issue can be simply solved by the double VP correction.

Resonant channel $e^+e^-\to\psi\to\mu^+\mu^-$ has two independent virtual photons, one is between $e^+e^-$ and $\psi$, and another is between $\psi$ and $\mu^+\mu^-$. According to the Feynman rule and ISR correction principle, each independent virtual photon propagator will be modified by a single VP correction factor, and the two VP factors cannot be combined into one. A Feynman diagram with time reversal symmetry can be plotted as Fig.~\ref{doublelevpamplitude}.

\begin{center}
\includegraphics[width=8.6cm]{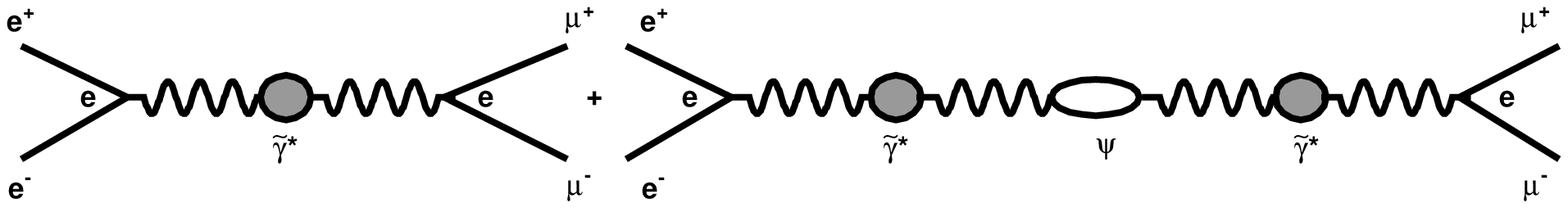}
\figcaption{\label{doublelevpamplitude} Feynman diagram with double VP correction.}
\end{center}

\begin{figure*}[t]
\centering
\includegraphics[width=6.21cm]{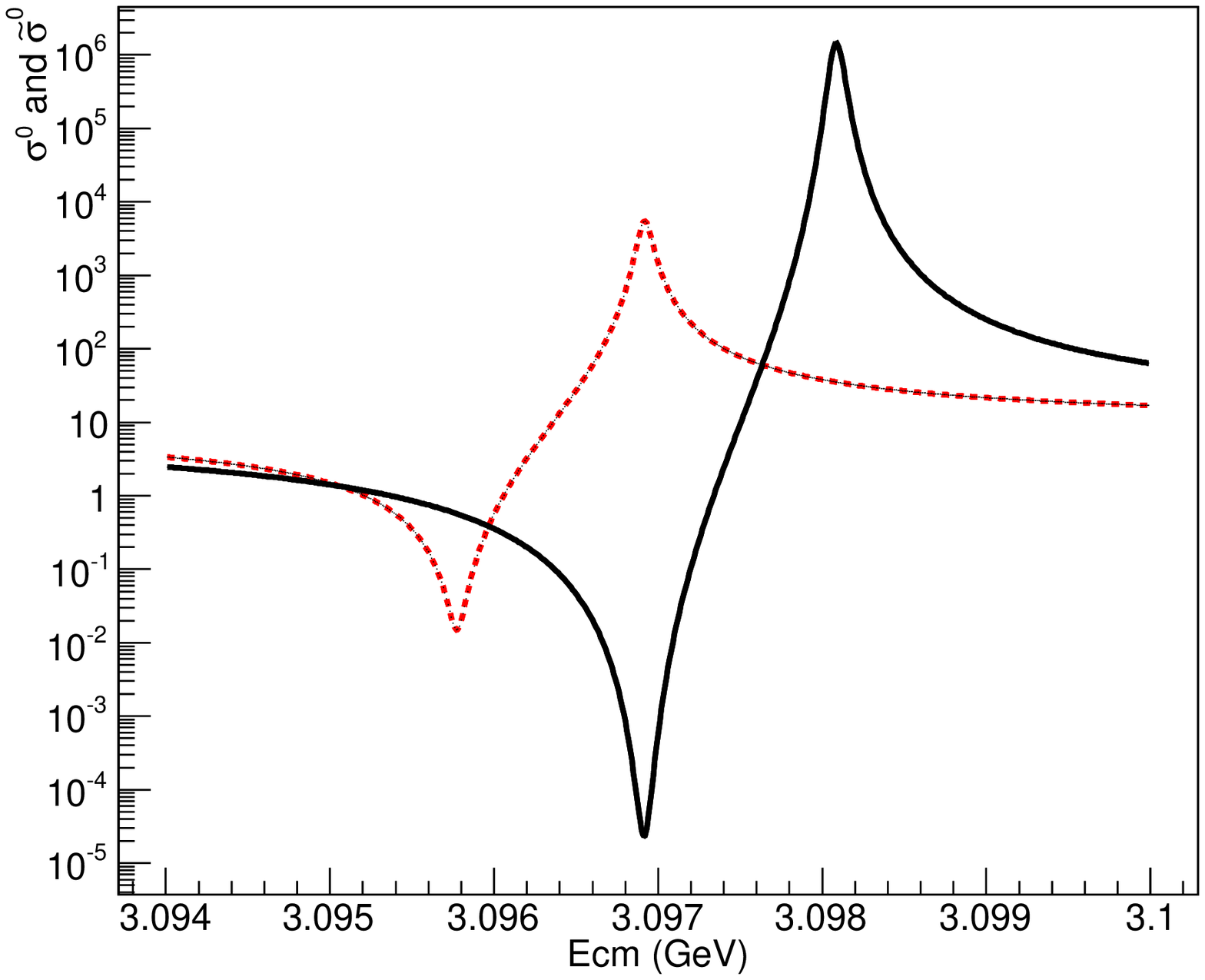}
\includegraphics[width=6.21cm]{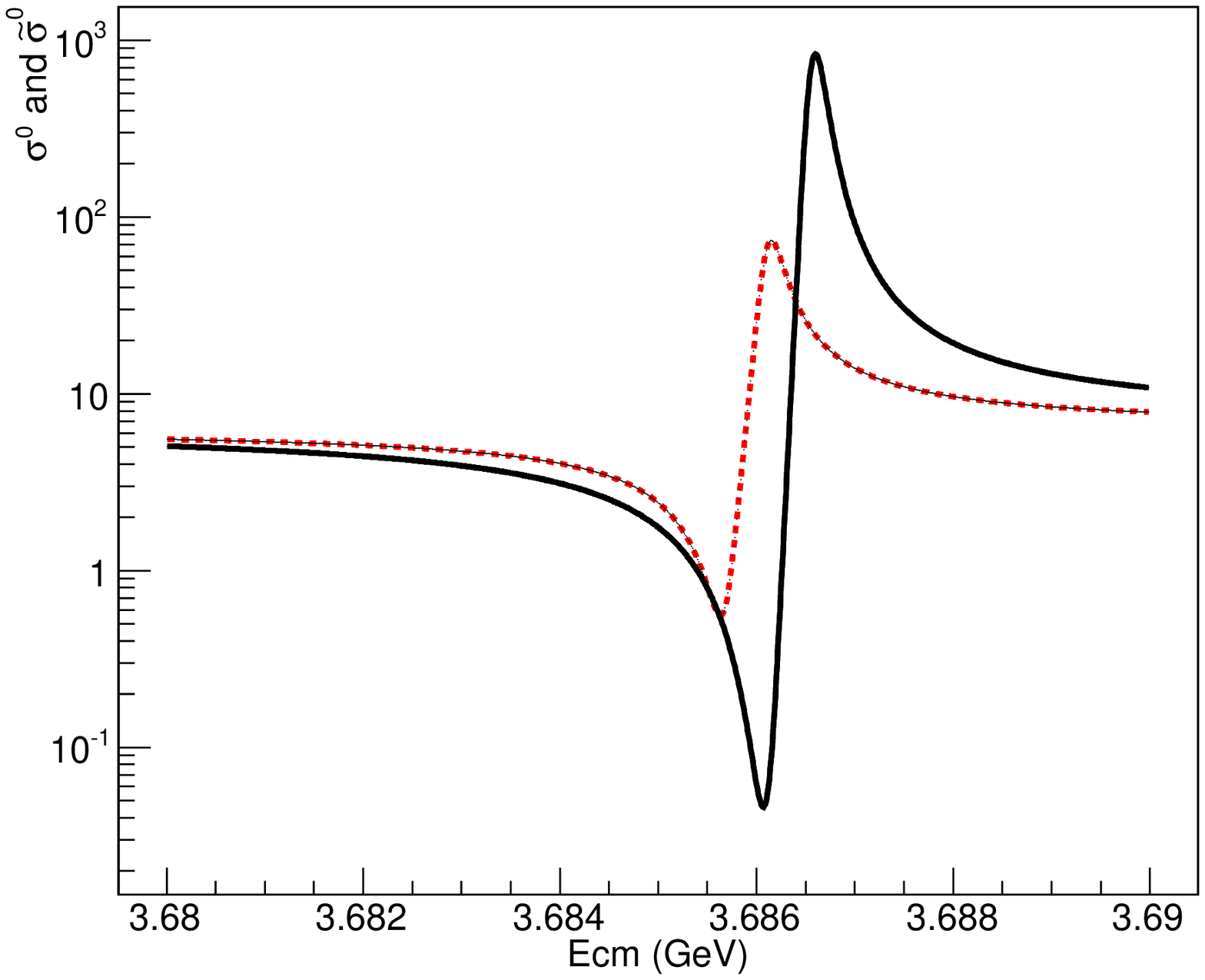}
\figcaption{\label{eemmborncrxtdbvp}(color online) Line-shape of $\sigma^0(s)$ (dashed line) and $\tilde{\sigma}^0(s)$ expressed in Eq.~(\ref{doublevpamplitudeformula}) and Eq.~(\ref{origbornduoblevpcor}) (solid line) in the vicinity of $J/\psi$ (left) and $\psi(3686)$ (right).}
\end{figure*}

The coherent amplitude for the Feynman diagram, as shown in Fig.~\ref{doublelevpamplitude}, after the contraction of the Lorentz indices of the virtual photons $\gamma^\ast$ and intermediary vector meson $\psi$, can be written as:
\begin{equation}
\tilde{{\cal A}}_{\rm eff}\sim\frac{1}{1-\hat\Pi(s)}+\frac{1}{1-\hat\Pi(s)}\frac{Fr{\rm e}^{{\rm i}\delta}}{\Delta+ir}\frac{1}{1-\hat\Pi(s)},\label{doublevpamplitudeformula}
\end{equation}
and the corresponding cross-section is:
\begin{equation}
\tilde{\sigma}^0(s)=\frac{4\pi\alpha^2}{3s}|\tilde{{\cal A}}_{\rm eff}|^2.\label{origbornduoblevpcor}
\end{equation}
Figure~\ref{eemmborncrxtdbvp} presents the line-shape comparison of $\sigma^0(s)$ expressed in Eq.~(\ref{totalborncrxt}) and $\tilde{\sigma}^0(s)$ in Eq.~(\ref{origbornduoblevpcor}).

Comparing Figs.~\ref{eemmborncrxtsgvp} and ~\ref{eemmborncrxtdbvp}, the single and double VP correction lead to different line-shapes for the cross-section. This issue will yield different results when extracting the resonant parameters from experimental data.

The Feynman diagram in Fig.~\ref{doublelevpamplitude} with double VP correction can be replotted equivalently as Fig.~\ref{doublelevpamplitudeeq}, which is symmetrical for the two time-reversal leptonic processes:
\begin{equation}
e^+e^-~~\rightleftarrows~~\gamma^\ast/\psi~~\rightleftarrows~~\mu^+\mu^-.
\end{equation}
The tree-level Feynman diagrams in Fig.~\ref{treeamplitude} and double VP-corrected equivalent diagram in Fig.~\ref{doublelevpamplitudeeq} have the same topology, but the coupling vertexes possess different coupling strengths $e$ and $e(s)$, respectively.

\begin{center}
\includegraphics[width=10.8cm]{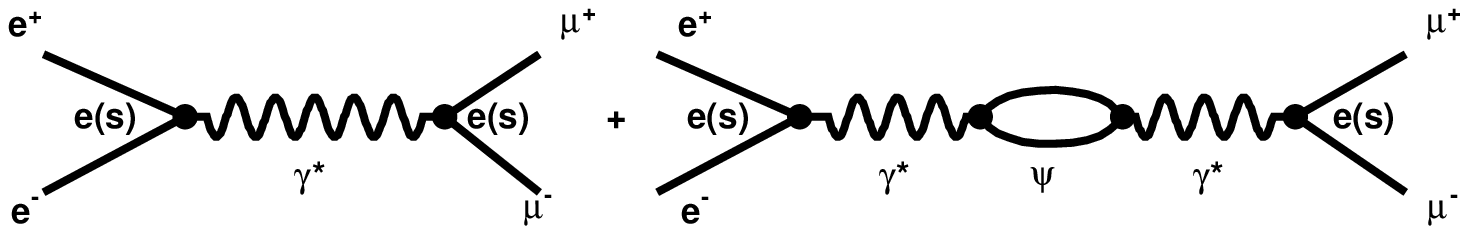}
\figcaption{\label{doublelevpamplitudeeq} Equivalent Feynman diagram with double VP correction; the black spots represent effective charge $e(s)$ defined by Eq.~(\ref{runninge}).}
\end{center}

\begin{figure*}[t]
\centering
\includegraphics[width=6.2cm]{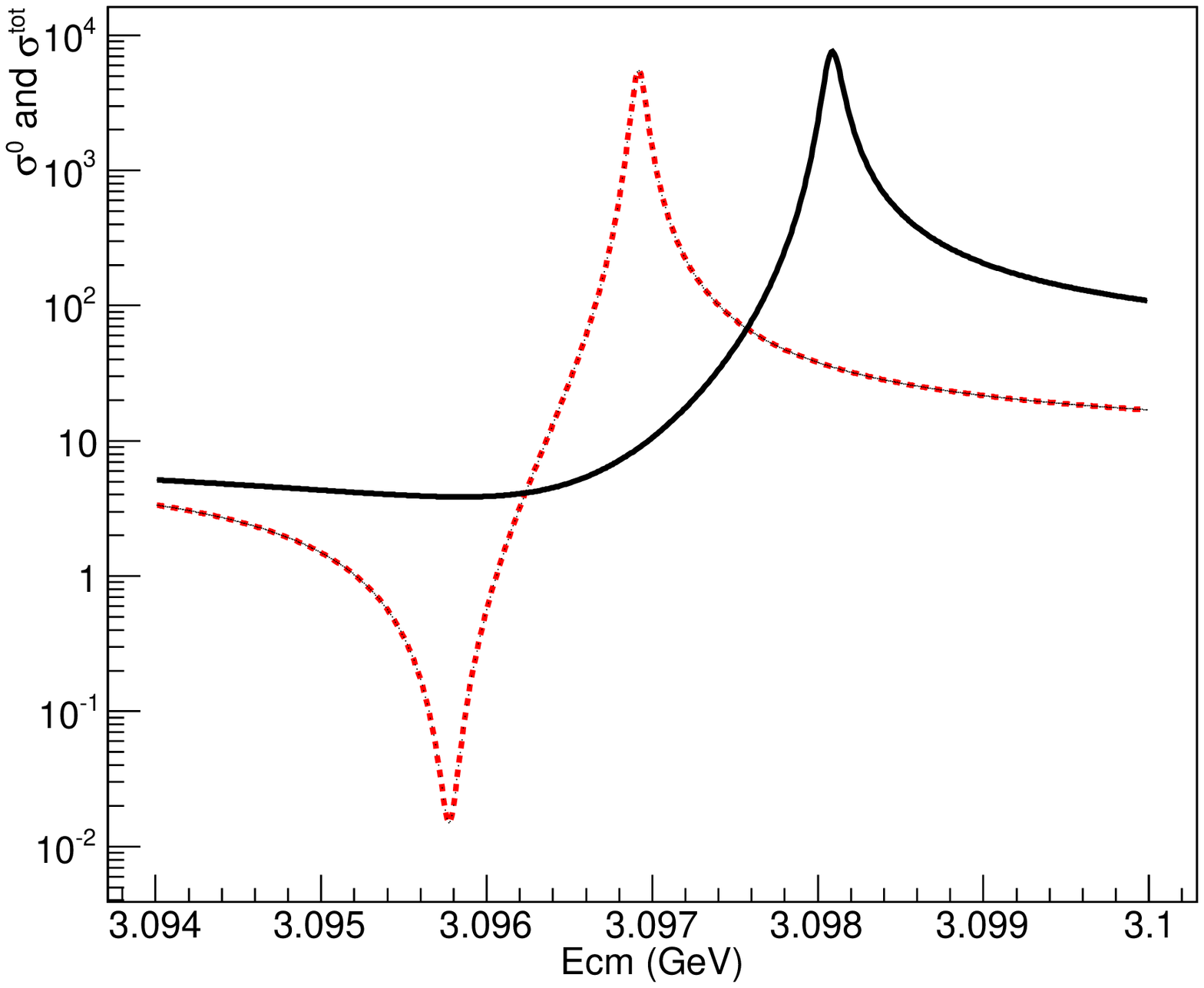}
\includegraphics[width=6.2cm]{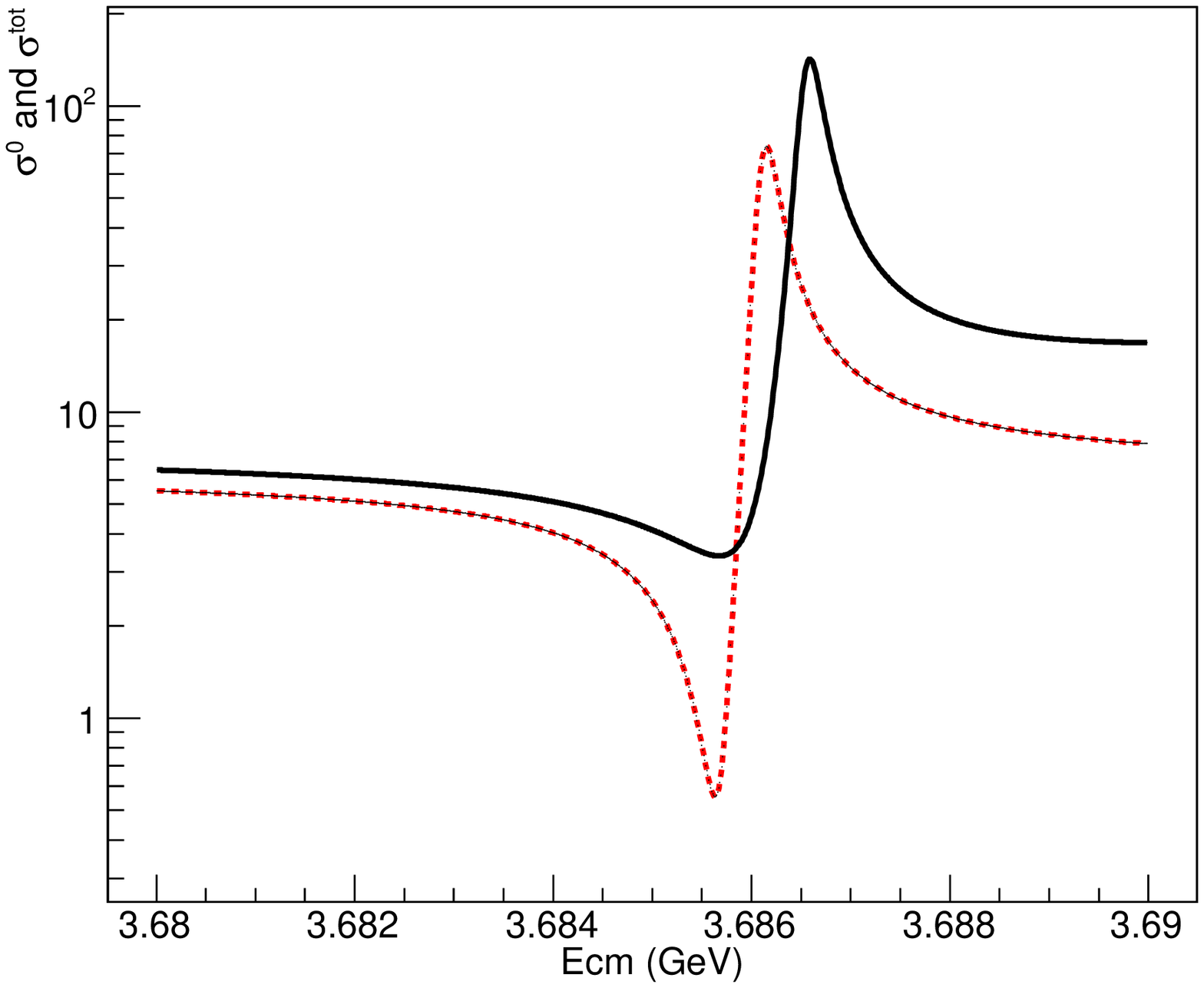}
\figcaption{\label{eemmcrborncrxtsgvp}(color online) Line-shapes of $\sigma^0(s)$ (dashed line) and $\sigma^{\rm tot}(s)$ for single VP correction (solid line) in the vicinity of $J/\psi$ (left) and $\psi(3686)$ (right).}
\end{figure*}

\begin{figure*}[t]
\centering
\includegraphics[width=6.2cm]{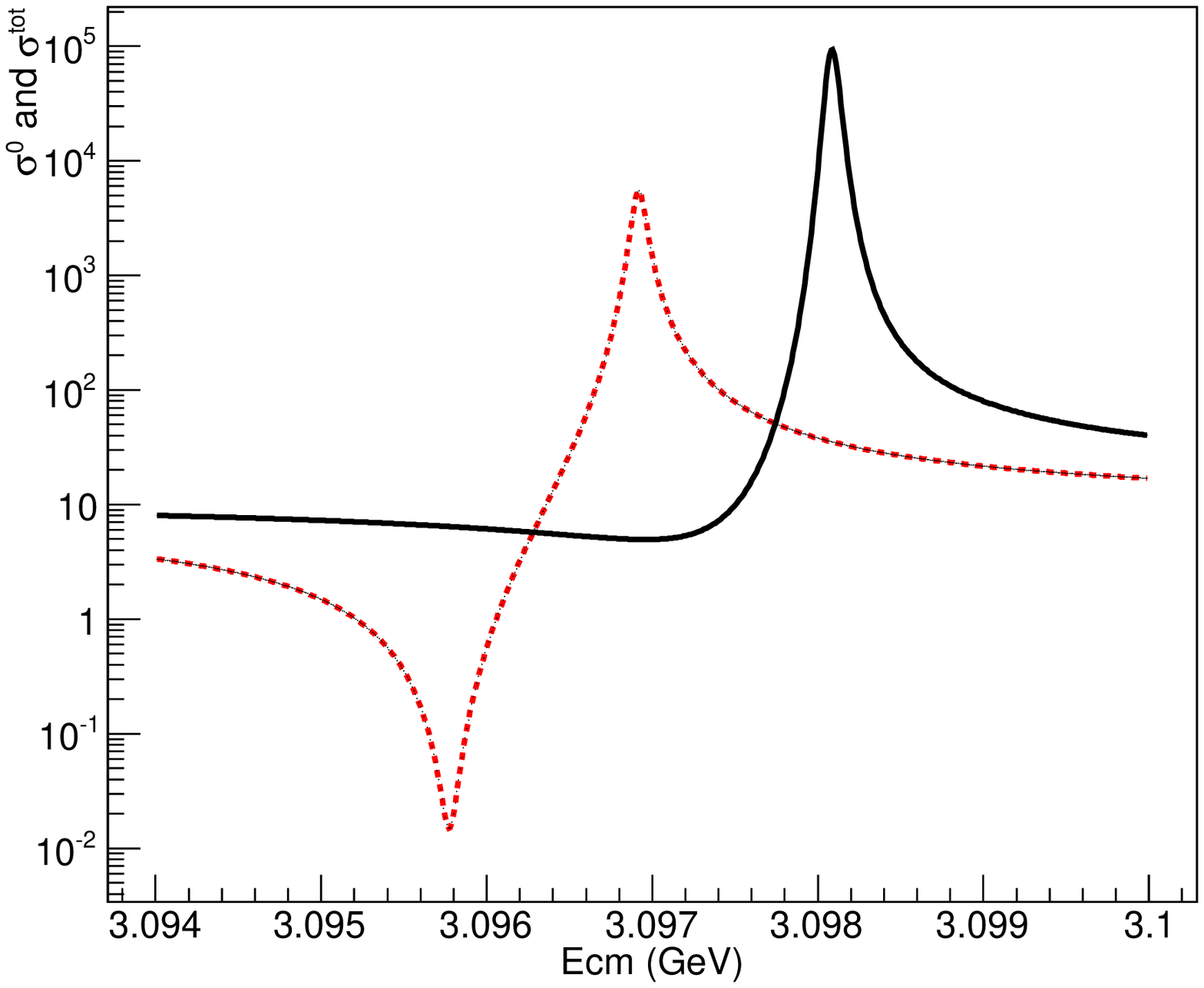}
\includegraphics[width=6.2cm]{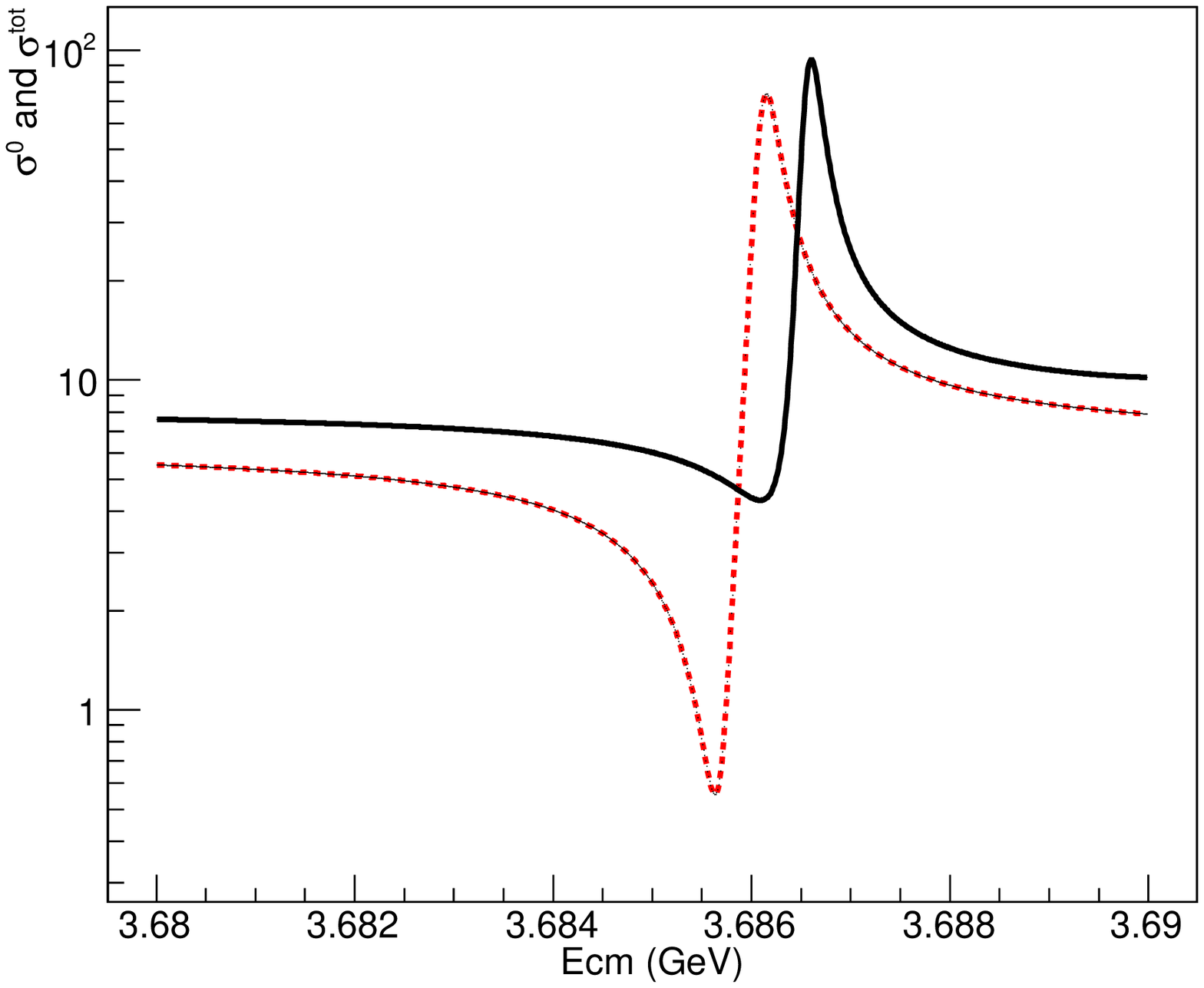}
\figcaption{\label{eemmcrborncrxtdbvp}(color online) Line-shapes of $\sigma^0(s)$ (dashed line) and $\sigma^{\rm tot}(s)$ for double VP correction (solid line) in the vicinity of $J/\psi$ (left) and $\psi(3686)$ (right).}
\end{figure*}

\begin{figure*}[t]
\centering
\includegraphics[width=6.2cm]{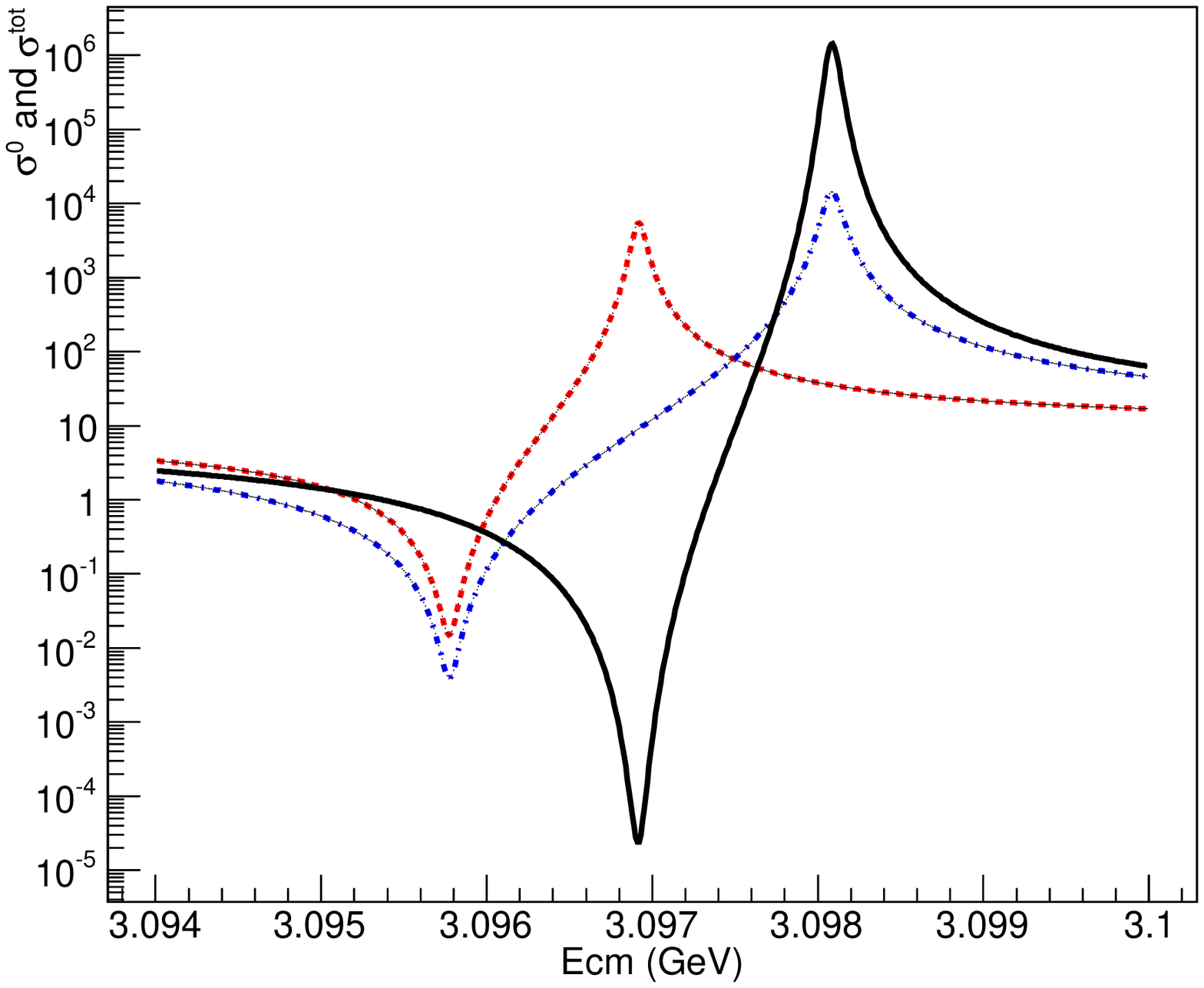}
\includegraphics[width=6.2cm]{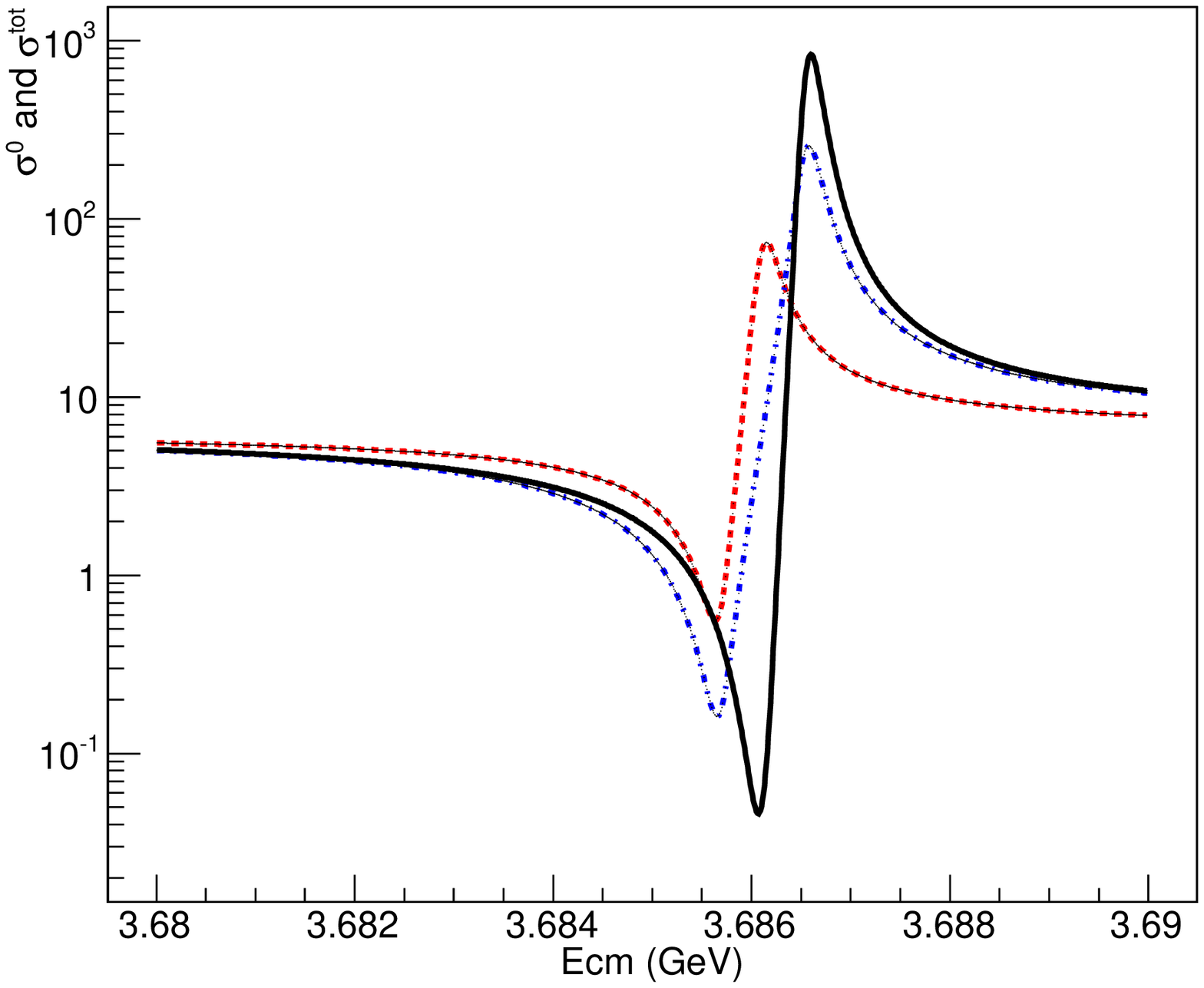}
\figcaption{\label{eemmcrborncrxtall}(color online) Line-shapes of $\sigma^0(s)$ (dashed line), single (dot-dashed line), and double (solid line) VP-modified $\tilde{\sigma}^0(s)$ in the vicinity of $J/\psi$ (left) and $\psi(3686)$ (right).}
\end{figure*}

\section{Total cross-section}

The Born cross-section corresponding to the tree-level Feynman diagram reflects the basic property of an elementary particle reaction process, which is interesting in physics. However, in experiments, the measured property is the total cross-section. In this section, the general form of the total cross-section for $e^+e^-\to\mu^+\mu^-$ is given first. Subsequently, the analytical expression of the total cross-section is deduced for the cases of single and double VP corrections, and they are compared numerically.

\subsection{General form}

In the Feynman diagram scheme, the total cross-section up to order ${\cal O}(\alpha^3)$ can be written as\cite{slac4160,slac5160}:
\begin{equation}
\sigma^{\rm tot}(s)=(1-x_m^\beta+\delta_{\rm vert})\tilde\sigma^0(s)+\int_{0}^{x_m}{\rm d}xH(x;s)\tilde\sigma^0(s'),\label{sigmatotfm}
\end{equation}
where $x\equiv E_\gamma/\sqrt{s}$ is the energy fraction carried by a Bremsstrahlung photon,
$x_m=1-4m_\mu^2/s$ is the maximum energy fraction of the radiative photon, $s'=(1-x)s$ is the effective square of the center-of-mass energy of the final $\mu^+\mu^-$ pair after radiation, $\delta_{\rm vert}$ is the vertex correction factor, and the radiative function is:
\begin{equation}
H(x;s)=\beta\frac{x^\beta}{x}\left(1-x+\frac{x^2}{2}\right),~~~~\beta=\frac{2\alpha}{\pi}\left(\ln\frac{s}{m_e^2}-1\right).\label{radiatorfm}
\end{equation}

In principle, the integral in Eq.~(\ref{sigmatotfm}) can be calculated using a numerical method. However, in the application for narrow resonances $J/\psi$ and $\psi(3686)$ scan experiment, the $e^\pm$ beam energy spread effect must be considered. The effect total cross-section that matches the experiment data is:
\begin{equation}
\sigma_{th}^{\rm tot}(s_0)=\int {\rm d} sG(s;s_0)\sigma^{\rm tot}(s),\label{energyspread}
\end{equation}
where $G(s;s_0)$ is the Gaussian function representing the energy spread distribution of the initial $e^\pm$ beams and $\sqrt{s_0}$ is the nominal center-of-energy of $e^\pm$. Eq.~(\ref{energyspread}) is a two-dimensional integral in variables $x$ and $s$. Integral Eq.~(\ref{energyspread}) contains Eq.~(\ref{sigmatotfm}) and the outer integral in $s$ about energy spread has to be calculated numerically. However, the inner integral in Eq.~(\ref{sigmatotfm}) of $x$ can be evaluated analytically. The analytical calculation in Eq.~(\ref{sigmatotfm}) can save much CPU time and achieve high numerical accuracy.

In the following sections, the analytical expression of integral Eq.~(\ref{sigmatotfm}) is deduced for the two cases of single and double VP corrections, and total cross-section $\sigma^{\rm tot}(s)$ is evaluated using the analytical results.

\subsection{Analytical calculation for single VP}

If the initial $e^\pm$ radiates a photon with energy fraction $x$, the notations in Eqs.~(\ref{deltadefold}) and ~(\ref{pires}) are changed:
\begin{equation}
\Delta \Rightarrow \Delta(x)=(1-x)t-1,
\end{equation}
\begin{equation}
\Pi_{\rm res}(s)\Rightarrow\Pi_{\rm res}(x;s)=h\frac{1-x}{\Delta(x)+ir}.
\end{equation}
The Born cross-section with VP correction is:
\begin{equation}
\sigma^0(s)\Rightarrow\tilde{\sigma}^0(x;s)=\frac{4\pi\alpha^2}{3s}\frac{1}{1-x}\cdot\frac{U(x)}{V(x)},
\end{equation}
where the quadratic polynomials have the forms:
\begin{eqnarray}
U(x)&=&u_2x^2+u_1x+u_0,\\
V(x)&=&v_2x^2+v_1x+v_0.
\end{eqnarray}
The integrand in Eq.~(\ref{sigmatotfm}) has the following polynomial form:
\begin{align}
H(x)\tilde{\sigma}^0(x;s)=\frac{4\pi\alpha^2}{3s}\beta\frac{x^\beta}{x}\left[\frac{1}{1-x}\sum_{n=0}^4w_nx^n
                         +\frac{1}{V(x)}\sum_{n=0}^5d_nx^n\right],
\end{align}
where coefficients $u_i$, $v_i$, $w_n$, and $d_n$ are the combinations of known constants and resonant parameters. The integral of Eq.~(\ref{sigmatotfm}) can be performed analytically. The results of the analytical integrals of $\sigma^{\rm tot}(s)$ are shown in Fig.~\ref{eemmcrborncrxtsgvp}, and the line-shape of $\sigma^0(s)$ is plotted to exhibit the effect of the ISR correction.

\subsection{Analytical calculation for double VP}

The integrand of Eq.~(\ref{sigmatotfm}) for the double VP correction can be expressed as the following elementary function:
\begin{eqnarray}\label{integranddbvpint}
H(x)\tilde{\sigma}^0(x;s)&=&\dfrac{4\pi\alpha^2}{3s}\beta\frac{x^\beta}{x}\bigg[\dfrac{1}{1-x}\sum_{n=0}^4p_nx^n+\dfrac{1}{V(x)}\sum_{n=0}^5q_nx^n+\dfrac{1}{V^2(x)}\sum_{n=0}^5r_nx^n\bigg],~~~~
\end{eqnarray}
where coefficients $p_n$, $q_n$, and $r_n$ are the combinations of known constants and resonant parameters. The integral of Eq.~(\ref{sigmatotfm}) can be performed analytically, and the analytical results are displayed in Fig.~\ref{eemmcrborncrxtdbvp}.

\section{Discussions}

This work discusses two issues: (1) treating the VP correction of the $\gamma^\ast$ channel and $\psi$ channel by a natural and consistent scheme; (2) comparing the cross-sections of $e^+e^-\to\gamma^\ast/\psi\to\mu^+\mu^-$ evaluated by the single and double VP corrections schemes.

The tree-level Feynman diagram in Fig.~\ref{treeamplitude} for $e^+e^-\to\gamma^\ast/\psi\to\mu^+\mu^-$ is the coherent summation of the $\gamma^\ast$ channel and $\psi$ channel. The VP-modified Born cross-section is given in Eq.~(\ref{origbornduoblevpcor}), the $\gamma^\ast$ channel is modified by a single VP factor, and the $\psi$ channel is modified by double VP factors.

Figure~\ref{eemmcrborncrxtall} exhibits the comparison of original Born cross-section $\sigma^0(s)$ and single and double VP-modified Born cross-sections $\tilde{\sigma}^0(s)$ in the vicinity of $J/\psi$ and $\psi(3686)$. The line-shapes of $\tilde{\sigma}^0(s)$ for the single and double VP corrections are significantly different.

Reference \cite{agshamov} discusses the VP-modified Born cross-section of process $e^+e^-\to\mu^+\mu^-$, where the tree-level Feynman diagram is only a continuum $\gamma^\ast$ channel and there is no resonant $\psi$ channel. In fact, this is the case discussed in section 5.1 in this paper. The VP-modified Born cross-section in reference \cite{agshamov} is same as expressed in Eq.~(\ref{convp}) in our paper. Eq.~(\ref{convp}) is a very concise and natural expression, and it is easy to understand in physics. Reference \cite{agshamov} made a skillful mathematic identical transformation to VP correction, where the full factor of $1/(1-\hat\Pi)$ was divided to two terms: the term with $1/(1-\Pi_0)$ explained as the continuum amplitude, and term $\tilde{\Pi}_{\rm res}/(1-{\Pi}_0)^2$ as the resonant amplitude. In this explanation, only non-resonant component $\Pi_0$ is viewed as the VP correction factor, whereas resonant component $\tilde{\Pi}_{\rm res}$ is viewed as the resonant amplitude. Thus, the original one-continuum channel is transformed into two channels, which implies that a pure identical transformation in mathematics leads to a new physics picture. Resonant amplitude $\tilde{\Pi}_{\rm res}$ contains non-resonant components $\Pi_0$ of $\hat\Pi$ in the following form:
\begin{equation}
\tilde{\Pi}_{\rm res}(s)=\frac{3\varGamma_e}{\alpha}\frac{s}{M}\frac{1}{s-\tilde{M}^2+i\tilde{M}\tilde{\varGamma}},\label{tildepires}
\end{equation}
where mass $\tilde{M}$ and width $\tilde{\varGamma}$ are called dressed values:
\begin{equation}
\tilde{M}^2=M^2+\frac{3\varGamma_e}{\alpha}\frac{s}{M}{\rm Re}\frac{1}{1-\Pi_0},\label{tildem}
\end{equation}
\begin{equation}
\tilde{M}\tilde{\varGamma}=M\varGamma-\frac{3\varGamma_e}{\alpha}\frac{s}{M}{\rm Im}\frac{1}{1-\Pi_0}.\label{tildemgamma}
\end{equation}

Therefore, the value of $\varGamma_e^{ex}$ defined with convention Eq.~(\ref{gammaekedr}) cannot be adopted all alone because $\Pi_0$ is only a partial VP correction and not the full one, $\hat\Pi$. In this case, $\varGamma_e^{ex}$ must be used together with $\tilde{M}$ and $\tilde{\varGamma}$ for completeness and consistency. It is noticed that only $\varGamma_e$ is present in initial state $e^+e^-$ in the numerator of Eq.~(\ref{tildepires}) and that there is no $\varGamma_f$ for the appointed final state, $\mu^+\mu^-$. If $\tilde{\Pi}_{\rm res}$ can be interpreted as the resonant amplitude of $e^+e^-\to\psi\to\mu^+\mu^-$, why it cannot be for the other final states, such as $e^+e^-$, $\tau^+\tau^-$ or hadrons? In fact, the true resonant amplitude is written in the Breit--Wigner form in Eq.~(\ref{nonrrlativisticbwam}). The VP effect is the quantum fluctuation of vacuum, and it does not refer to any final state. Convention Eq.~(\ref{gammaekedr}) and the explanation in \cite{agshamov} convert a simple and clear problem as a complex and an obscure one. However, the convention in Eq.~(\ref{gammaetsai}) is clear and natural.

The bare resonant parameters $(M,\varGamma,\varGamma_e,\delta)$ are the basic quantities in the Breit--Wigner formula, and they characterize the main properties of a resonance. The values of these parameters can be estimated from phenomenological potential models \cite{eichten,wolfgang}. However, their accurate values have to be measured by fitting the experimental data.

Generally, the cross-section directly measured in experiments is the total cross-section, which includes all the radiative effects. To extract the bare resonant parameters from the measured cross-section correctly, an appropriate treatment of the ISR correction is crucial.

As seen in the previous sections, the value of the total cross-section, $\sigma_{th}^{\rm tot}(s)$, depends on the VP correction scheme, and it is also the function of the resonant parameters. ISR correction factor $1+\delta$ is a theoretical quantity defined in Eq.~(\ref{isrfactordef}), and it affects the Born cross-section according to Eq.~(\ref{croxtborndef}).

The values of the resonant parameters of $J/\psi$ and $\psi(3686)$ can be extracted by fitting the measured cross-section in the line-shape scan experiment based on the least square method:
\begin{equation}
\chi^2=\sum_{i=1}^n\frac{[\sigma_{ex}^{\rm tot}(s_i)-\sigma_{th}^{\rm tot}(s_i)]^2}{\Delta_i^2},\label{fitting}
\end{equation}
where $\sigma_{ex}^{\rm tot}$ can be measured using Eq.~(\ref{croxttotdef}) and $\Delta_i$ is the uncertainty of $\sigma_{ex}^{\rm tot}(s_i)$ at energy point $s_i$. The optimized values of $(M,\varGamma,\varGamma_e,\delta)$ correspond to the optimized minimum of $\chi^2$.

When the value of $\varGamma_e$ is extracted, one may obtain $\varGamma_e^{ex}$ by any convention, but it is not necessary in physics and nor in experiments. $\varGamma_e$ connects to original radial wave function $R(0)$ of $c\bar{c}$ bound state $\psi$ according to Eq.~(\ref{gammaemodel}). The value of $\varGamma_e$ can deduce the value of $R(0)$ and can test potential models. $\varGamma_e$ can be used to calculate the correct ISR factor in the $R$ measurement.

It is expected that if the values of the resonant parameters $(M,\varGamma,\varGamma_e,\delta)$ are extracted using the scheme proposed in this paper, the results will not be the same as in previous measurements. Therefore, which scheme is reasonable should be determined by experiments and further studies.


\vspace{-2.5mm} \centerline{\rule{80mm}{0.1pt}} \vspace{1mm}



\clearpage

\end{CJK*}
\end{document}